\documentclass[12pt]{emulateapj}
\slugcomment{{\sc Accepted for publication in The Astrophysical Journal:} August 15, 2011}

\def\etl{et al.}

\def\niilam{[\ion{N}{2}]~$\lambda6584$}

\def\oiiilam{[\ion{O}{3}]~$\lambda5007$}

\newcommand{\loiiinocor}{\ensuremath{L_{{\rm [OIII]\lambda}5007}}}
\newcommand{\loiii}{\ensuremath{L_{{\rm [OIII]\lambda}5007, {\rm obs}}}}
\newcommand{\loiiicor}{\ensuremath{L_{{\rm [OIII]\lambda}5007, {\rm cor}}}}
\newcommand{\lmaser}{\ensuremath{L_{\rm H2O}}}
\newcommand{\mbh}{\ensuremath{M_{\rm BH}}}
\newcommand{\lagn}{\ensuremath{L_{\rm AGN}}}
\newcommand{\lfir}{\ensuremath{L_{\rm FIR}}}
\newcommand{\lxray}{\ensuremath{L_{2-10}}}
\newcommand{\Rcr}{\ensuremath{R_{\rm cr}}}

\newcommand{\mb}{\ensuremath{M_{B}}}
\newcommand{\omegam}{\ensuremath{\Omega_{\rm m}}}
\newcommand{\omegal}{\ensuremath{\Omega_{\Lambda}}}

\newcommand{\hh}{\ensuremath{H_{0}}}
\newcommand{\kms}{\ensuremath{{\rm km~s}^{-1}}}
\newcommand{\mpc}{\ensuremath{{\rm Mpc}^{-1}}}
\newcommand{\ergscm}{\ensuremath{{\rm erg~s}^{-1}{\rm~cm}^{-2}}}
\newcommand{\ergs}{\ensuremath{{\rm erg~s}^{-1}}}

\newcommand{\Lsun}{\ensuremath{L_{\odot}}}

\newcommand{\Hb}{\ensuremath{{\rm H}\beta}}
\newcommand{\Ha}{\ensuremath{{\rm H}\alpha}}

\newcommand{\latin}[1]{{#1}}

\newcommand{\eg}{\latin{e.g.}}

\usepackage{natbib}

\begin{document}
\shorttitle{Hosts of Water Masers}
\shortauthors{Zhu \etl}
\title {Optical Properties of Host Galaxies of Extragalactic Nuclear Water Masers}

\author{
 Guangtun Zhu\altaffilmark{1},
 Ingyin Zaw\altaffilmark{2, 1}, 
 Michael R. Blanton\altaffilmark{1}, and
 Lincoln J. Greenhill\altaffilmark{3}
} 
\altaffiltext{1}{Center for Cosmology and Particle Physics, Department of Physics, 
New York University, 4 Washington Place, New York, NY 10003, 
gz323@nyu.edu}
\altaffiltext{2}{New York University Abu Dhabi, P.O. Box 903, New York, NY 10276}
\altaffiltext{3}{Harvard-Smithsonian Center for Astrophysics, 60 Garden Street, 
Cambridge, MA 02138}


\begin{abstract}
We study the optical properties of the host galaxies of nuclear $22$
GHz ($\lambda=1.35$ cm) water masers. To do so, we cross-match the
galaxy sample surveyed for water maser emission ($123$ detections and
$3806$ non-detections) with the SDSS low-redshift galaxy sample ($z<0.05$).
Out of 1636 galaxies with SDSS photometry, we identify $48$ detections;
out of the 1063 galaxies that also have SDSS spectroscopy, we identify
33 detections.  We find that maser detection rate is higher at higher
optical luminosity ($M_B$), larger velocity dispersion ($\sigma$), 
and higher \oiiilam{} luminosity, 
with \oiiilam{} being the dominant factor.
These detection rates are essentially the result of the correlations of
isotropic maser luminosity with all three of these variables.  
These correlations are natural if maser strength increases with central
black hole mass and the level of AGN activity.  We also find that the 
detection rate is higher in galaxies with higher extinction.
Based on these results, we propose that maser surveys seeking to 
efficiently find masers should rank AGN targets by 
extinction-corrected \oiiilam{} flux when available. 
This prioritization would improve maser detection efficiency, from an
overall $\sim3\%$ without pre-selection to $\sim16\%$ for the
strongest intrinsic \oiiilam{} emitters, by a factor of $\sim 5$.
\end{abstract}

\keywords{galaxies: active ---
  galaxies: nuclei  ---
  galaxies: Seyfert --- 
  radio lines: galaxies ---
  masers}

\section {Introduction}

Water maser emission at $22$ GHz ($\lambda=1.35$ cm) is currently the only 
tracer of warm dense molecular gas in the inner parsec of active 
galaxy nuclei (AGNs) and has been detected to date in more than $100$ AGNs
\citep[\eg,][]{braatz96a, henkel05a, kondratko06a, braatz08a, greenhill08a}.  
Some of these masers are associated with rotating, highly inclined disk 
structures close to the central engines (``disk masers'') and have been used for 
a broad variety of astrophysical studies, including the mass estimation of 
supermassive black holes, the mapping of accretion disks, and the determination 
of geometric distances \citep[\eg,][]{miyoshi95a, greenhill97a,
greenhill97b, ishihara01a, greenhill03a, braatz10a, kuo10a}. 

Nuclear water masers have been claimed 
to be associated with Seyfert 2 or low-ionization nuclear emission-line
region (LINER) systems \citep[\eg,][]{braatz97a, kondratko06a}. It is
also plausible that AGNs which host masers are more likely associated
with high X-ray obscuring columns ($N_H$) than those without maser
detections \citep[\eg,][]{braatz97a, madejski06a, zhang06a,
  greenhill08a, zhang10a}.  There also appear to be correlations of
isotropic maser luminosity with the X-ray luminosity
\citep[][]{kondratko06a} and the far-infrared (FIR) luminosity
\citep[][]{henkel05a} of the host AGNs, though the underlying
mechanisms are not clear.  As additional words of caution in
interpreting these correlations, the inferred X-ray luminosities are
subject to large uncertainties owing to high columns, and the true
(i.e., beamed) maser luminosities are unknown in most cases.

The overall detection rate of nuclear water masers is only $\sim3\%$.
Even if AGNs with higher X-ray luminosity and/or higher obscuring
column more likely host masers, there is no existing large sample of
AGNs with X-ray data available for target selection. However, if
masers are preferentially found in galaxies with certain optical
properties, we can improve maser detection efficiency by selecting
galaxies with these properties as targets from existing large galaxy
surveys, such as the Sloan Digital Sky Survey
\citep[SDSS;][]{york00a}, the 2dF Galaxy Redshift Survey
\citep[2dFGRS;][]{colless01a}, and the 6dF Galaxy Survey
\citep[6dFGS;][]{jones04a}.

The goal of this work is to systematically investigate 
the optical properties of maser host galaxies.  We cross-match the SDSS 
low-redshift galaxy catalog with the complete galaxy sample surveyed for maser emission.
We find that maser detection rate is higher at higher optical 
luminosity, larger velocity dispersion, higher \oiiilam{} luminosity,
and higher extinction. 
We present these results in Section \ref{sec:efficiency}.  
In Section \ref{sec:relations}, we suggest that a plausible explanation of 
these results is that maser strength is correlated with the central black 
hole mass and the AGN activity of the host galaxies.  
In Section \ref{sec:strategy}, we suggest that maser 
surveys rank AGN targets by extinction-corrected \oiiilam{} flux, which should 
greatly improve the detection efficiency. 
We adopt a $\Lambda \mathrm{CDM}$ cosmology with $\omegam=0.3$, 
$\omegal=0.7$ and $\hh=70~\kms~\mpc$.

\section{Detection Efficiency}\label{sec:efficiency}

\begin{figure}
\epsscale{1.1}
\plotone{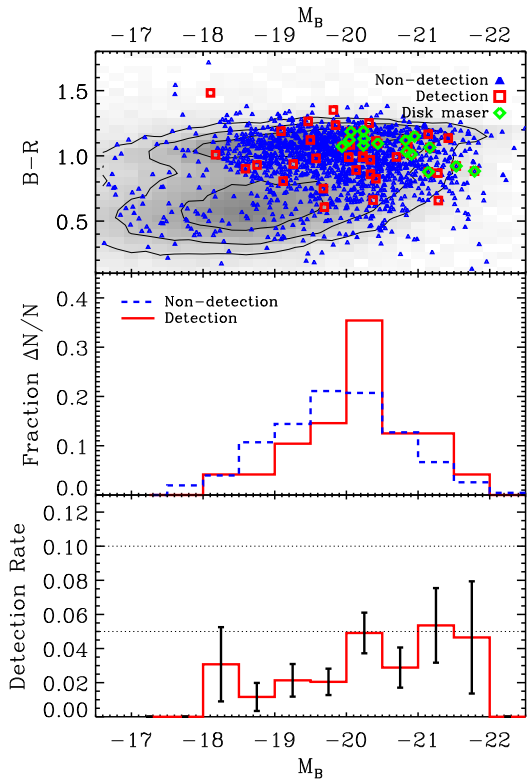}
\caption{
{\it Top panel}: Color-magnitude diagram. For comparison, we show the 
distribution of the whole low-z photometric sample as the gray scale.
The contours enclose $40\%$, $80\%$, and $90\%$ of the sample.
Blue triangles represent non-detections and red squares indicate maser detections.
We also show the disk masers with green diamonds but we cannot draw
robust conclusions because of the small sample size.
{\it Middle panel}: Distribution of \mb.  We show the ratio of the 
number of galaxies per $0.5$ mag to the size of each sample.
{\it Bottom panel}: Detection rate as a function of \mb.
To guide the eye, we show two horizontal dotted lines at $5\%$ and
$10\%$.  The error bars represent Poisson errors. 
The detection rate appears to be higher at higher luminosity.
}
\label{fig:mbeff}
\end{figure}

\subsection{Data}
\subsubsection{The complete galaxy sample surveyed for maser emission }

\begin{figure}
\epsscale{1.1}
\plotone{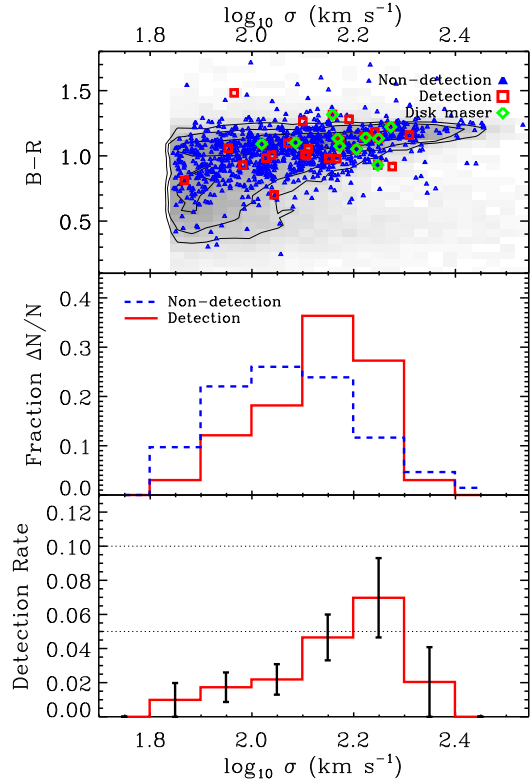}
\caption{
Similar to Figure \ref{fig:mbeff}, but with velocity dispersion ($\sigma$).
Note $\sigma$ lower than $70~\kms$ is not reliable due to the instrumental 
resolution of the SDSS spectrograph; we do not consider
galaxies below that limit.  The detection rate is higher at larger $\sigma$.
}
\label{fig:sigmaeff}
\end{figure}

To construct a complete sample of galaxies surveyed to date for maser emission, 
we combine the catalogs (as of December 1, 2010) maintained on the website
of the Megamaser Cosmology Project 
(MCP\footnote{\tt https://safe.nrao.edu/wiki/bin/view/Main/MegamaserCosmologyProject})
and that of the Hubble Constant Maser Experiment 
(HoME\footnote{\tt https://www.cfa.harvard.edu/$\sim$lincoln/demo/HoME/index.html}).
For maser detections, we use the MCP catalog that is complete. 
We however exclude those masers known to be associated with 
star-forming regions (IC 10, M 33, IC 342, M 82, NGC 253, NGC 3359, NGC 3556, NGC 2146, 
He 2-10, NGC 4038/39, NGC 4214, NGC 5253), as noted in the either HoME or MCP 
catalogs. Although it is not labeled in either catalog, we also exclude 
NGC 4194, since it is an ongoing merger and has strong star formation in the 
center \citep[\eg,][]{balzano83a} and the detected maser may not be associated with its nucleus. 
There are $123$ detections in total and we list them in Table \ref{table:data}.
Among these, at least $41$ are probably associated with disk structures, 
as noted in the HoME catalog. The evidence is either 
from direct mapping of the emission distributions using Very Long Baseline 
Interferometry (VLBI) or inferred from spectroscopy 
\citep[\eg,][]{madejski06a,greenhill08a}.
For galaxies without successful maser detections (non-detections), we combine 
both catalogs from HoME and MCP to build the whole sample. 
After removing duplicates between the catalogs, we have $3806$ non-detections in total.
The overall detection rate is therefore about $3\%$.

\begin{figure*}
\epsscale{0.9}
\plotone{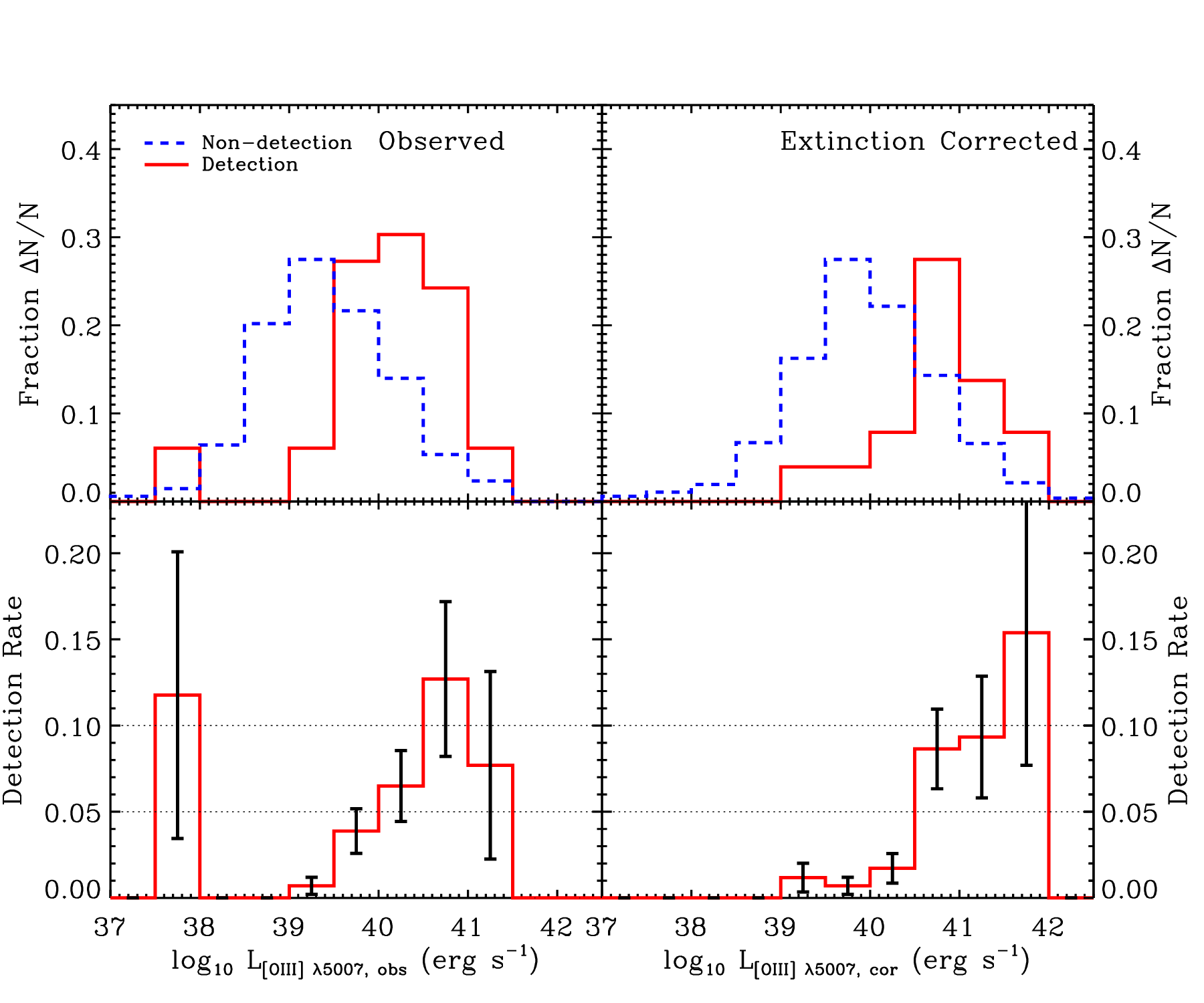}
\caption{
The left panels are similar to the lower two panels in Figure \ref{fig:mbeff} 
and \ref{fig:sigmaeff}, but with observed \oiiilam{} luminosity ($\loiii$). 
In the right panels, we correct the extinction to obtain
the intrinsic \oiiilam{} luminosity ($\loiiicor$) 
using $\loiiicor = \loiii/((\Ha/\Hb)/(\Ha/\Hb)_0)^{2.94}$, 
where $\Ha/\Hb$ is the observed Balmer decrement and 
we assume the intrinsic Balmer decrement $(\Ha/\Hb)_0=3$.
The detection rate  is higher at higher \oiiilam{} luminosity. 
The effect is stronger for \loiiicor, implying that detections
on average have higher extinction than non-detections.
}
\label{fig:oiiieff}
\end{figure*}

We note that this complete sample consists of galaxies surveyed using different 
telescopes with different detection sensitivities. The highest sensitivity 
comes from the Green Bank Telescope (GBT) survey \citep[\eg,][]{braatz04a}. 
The $1\sigma$ rms sensitivity of the GBT survey is $\sim3$ 
mJy per $24.4$ kHz ($\sim0.33~\kms$) channel \citep{braatz04a}. 
Assuming a characteristic maser linewidth of $10~\kms$, this corresponds to a 
$3\sigma$ maser flux limit of $0.1$ Jy $\kms$.

\subsubsection{The SDSS low-z galaxy sample}

To systematically study maser detection efficiency, we require a complete 
parent sample.  The SDSS survey has provided such a sample of galaxies 
with uniform imaging and spectroscopy.
For spectral properties, we use the measurements by the MPA-JHU 
group\footnote{\tt http://www.mpa-garching.mpg.de/SDSS/DR7/}
\citep[\eg,][]{tremonti04a}.
We use the latest version that corresponds to SDSS Data Release 7 
\citep[DR7,][]{abazajian09a}.
We choose to look at velocity dispersion ($\sigma$, in $\kms$) and \oiiilam{}
luminosity ($\loiiinocor$, in $\ergs$), since velocity dispersion 
is closely related to central black hole mass 
\citep[\eg,][]{ferrarese00a, gebhardt00a} 
and \oiiilam{} luminosity is well-correlated with AGN activity
\citep[\eg,][]{heckman05a}. It is well-known that \oiiilam{} can be 
severely obscured by material in the host galaxy 
\citep[\eg,][]{diamond09a}. We therefore calculate the intrinsic 
\oiiilam{} luminosity ($\loiiicor$) by correcting the observed $\loiii$ 
using the following formula \citep[\eg,][]{bassani99a}:
$\loiiicor = \loiii/((\Ha/\Hb)/(\Ha/\Hb)_0)^{2.94}$, 
where $\Ha/\Hb$ is the observed Balmer decrement and 
we assume the intrinsic Balmer decrement $(\Ha/\Hb)_0=3$. Because of 
the instrumental dispersion of the SDSS spectrograph,
velocity dispersion measurements smaller than $70~\kms$
are not reliable\footnote{\tt
http://www.sdss.org/dr7/algorithms/veldisp.html}; we thus
only consider galaxies with $\sigma>70~\kms$.  
Finally, since only four (4C $+05.19$, SDSS J0804+3607, Mrk 34, and 3C 403) 
out of the $123$ maser detections 
are farther than $z=0.05$ and all of them are not in the MPA-JHU catalog,
we limit the sample to low-redshift galaxies with $z<0.05$.
At the faint end, the flux limit of the SDSS spectroscopic survey is 
$r = 17.77$, which corresponds to $M_B \sim -18$ at $z = 0.05$. 

Due to the difficulty of automatic photometric processing of big
galaxies, the SDSS catalog is missing many nearby, bright galaxies,
even though they are contained within the SDSS imaging footprint. 
For photometry, we therefore use the low-z catalog ($z<0.05$) from
the NYU Value Added Galaxy Catalog
(NYU-VAGC\footnote{\tt http://sdss.physics.nyu.edu/vagc/}; 
\citealt{blanton05a}).
This low-z photometric catalog includes any low-redshift galaxies from the 
Third Reference Catalog of Bright Galaxies (RC3; 
\citealt{devaucouleurs91a, corwin94a}) for which we have $ugriz$ 
imaging from SDSS, but which are not in the SDSS catalog.
We use the latest version of this catalog that corresponds to SDSS 
Data Release 6 (DR6, \citealt{adelman06a}). 
We have compared the photometry of those galaxies in the SDSS catalog 
with that from DR7 and found they are very consistent, therefore using
DR6 for photometry should not introduce any bias.
We derive absolute magnitudes using the {\tt kcorrect} package
\citep[v4.1.4;][]{blanton07a}. 
For easier comparison with previous studies, we choose the $B$ band 
magnitude $M_B$ to indicate optical luminosity.
Note here the magnitude is the total magnitude for the whole galaxy.

\begin{figure}
\epsscale{1.0}
\plotone{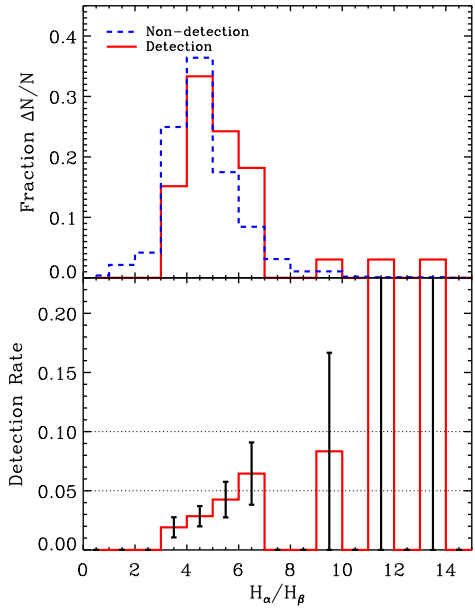}
\caption{
Similar to the lower two panels in Figure \ref{fig:mbeff} and \ref{fig:sigmaeff},
but with Balmer decrement $\Ha/\Hb$.
The detection rate  is higher at higher $\Ha/\Hb$.
}
\label{fig:hahbeff}
\end{figure}

\begin{figure}
\epsscale{1.1}
\plotone{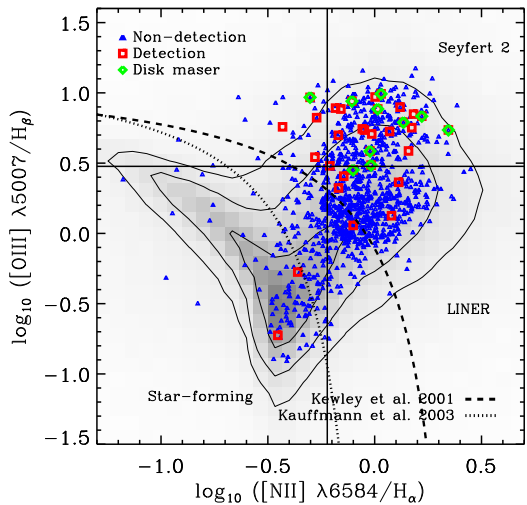}
\caption{
Emission line diagnostic diagram \citep[][BPT]{baldwin81a}.
For comparison, we show the distribution of the whole low-z 
spectroscopic sample as the gray scale.
The contours enclose $40\%$, $80\%$, and $90\%$ of the sample.
The dotted and dashed lines are the demarcation lines separating 
AGN and star-forming galaxies defined by 
\citet{kauffmann03a} and \citet{kewley01a}.
The solid vertical and horizontal lines at 
\niilam/\Ha=$0.6$ and \oiiilam/\Hb=$3.0$ are the conventional
separating lines for Seyfert 2 galaxies (above) and LINERs (below).
There are $25$ ($8$) detections above (below) the horizontal line,
compared to $296$ ($734$) non-detections.
}
\label{fig:bpt}
\end{figure}

We cross-match maser detections and non-detections with the SDSS low-z
sample, and identify $48$ detections ({\bf $15$} disk masers) and $1588$ 
non-detections with SDSS photometry, among which, $33$ detections ({\bf $10$} 
disk masers) and $1030$ non-detections have reliable spectral measurements, 
i.e., with $\sigma > 70~\kms$.
We note that the overall detection rate in this sample is $\sim3\%$, the same
as in the total sample, so using this sample should not introduce a bias in the 
analysis presented below.
We present measurements for detections in Table \ref{table:data}.
In the next subsection, we study the dependence of maser detection 
efficiency on optical luminosity and spectral properties using these 
samples.

\subsection{Results}
\subsubsection{All galaxies surveyed for water masers}

Figure \ref{fig:mbeff}, \ref{fig:sigmaeff}, and \ref{fig:oiiieff}
present maser detection efficiency as a function of $M_B$, 
$\sigma$ (in log scale), $\loiii$ (in log scale), 
and $\loiiicor$ (in log scale). In the top panels of Figure \ref{fig:mbeff} and
{\ref{fig:sigmaeff}, we show the detections (blue triangle) and 
non-detections (red square) in the color-magnitude/$\sigma$ diagram. 
For comparison, we also show the distribution of the whole SDSS low-z 
sample in gray scales and contours.  
Masers are concentrated in systems with larger $B-R$ and total luminosity,
but there are very few blue, low luminosity (presumably disk-dominated) galaxies
that have been surveyed for emission.
We also show the disk masers 
with green diamonds, but we cannot draw robust conclusions because 
of the small sample size. In the middle panels of Figure \ref{fig:mbeff} and 
\ref{fig:sigmaeff} and in the top panels of Figure \ref{fig:oiiieff}, we show 
the distribution of detections and non-detections. Performing 
Kolmogorov-Smirnov test yields P-values $0.01$, $0.02$, 
$2\times10^{-8}$, and $3\times10^{-9}$, respectively,
indicating a low probability that they are drawn from the same
distribution, particularly for $\loiiinocor$.
The bottom panels present the detection rate, which is apparently 
higher at brighter $M_B$, larger $\sigma$, and higher $\loiiinocor$.
The dependence of the detection efficiency on $\sigma$ appears stronger 
than that on $M_B$, while the dependence on $\loiiinocor$ is 
more striking than that on both $M_B$ and $\sigma$. 

Figure \ref{fig:oiiieff} also shows that maser detection efficiency depends 
more strongly on $\loiiicor$ than on $\loiii$, implying that detections could 
on average have higher extinction than non-detections. 
Previous studies show that AGNs which host masers are more likely associated
with high X-ray obscuring columns than those without maser detections 
\citep[\eg, ][]{greenhill08a, zhang10a}. We therefore investigate 
the detection efficiency as a function of the observed Balmer decrement 
in Figure \ref{fig:hahbeff}. In the top panel, a Kolmogorov-Smirnov test 
yields a P-value 0.009, indicating that detections and non-detections are likely drawn from 
different distributions. The lower panel shows the detection rate is 
indeed higher at higher extinction.

In Figure \ref{fig:bpt}, we show the emission line diagnostic diagram 
\citep[][BPT]{baldwin81a}. 
We show two widely-used demarcation criteria that separate AGNs (to the right) 
and star-forming galaxies (to the left), by \citet[][the dashed line]{kewley01a}
and by \citet[][the dotted line]{kauffmann03a}.  The vertical solid line at 
\niilam/\Ha$=0.6$ and the horizontal solid line at 
\oiiilam/\Hb$=3.0$ are the conventional demarcation lines for Seyfert 2
(above) and LINER-like (below) galaxies \citep[\eg,][]{veilleux87a}.
It is apparent that most surveys have mainly targeted AGNs.

Among AGNs, masers are clearly more often detected in Seyfert 2
galaxies rather than LINERs. For example, in Figure \ref{fig:bpt}, there are $25$ maser
detections associated with Seyfert 2 galaxies (out of $321$ in total, above 
the horizontal line), but only
eight associated with LINERs (out of $742$ in total, below the horizontal line).
It is yet unkown what fraction of LINERs are AGNs \citep[\eg, ][]{ho03a, sarzi10a} and 
we are not sure about the underlying mechanisms responsible 
for the preferred association of masers with Seyfert 2 galaxies over LINERs. 
However, if LINERs are low-luminosity counterparts of Seyfert 2 galaxies,
as some studies claim \citep[\eg,][]{ho03a}, then the low maser detection efficiency 
among LINERs could be simply a reflection of the $\loiiinocor$ dependence 
we found in Figure \ref{fig:oiiieff}, since the majority of LINERs 
without maser detections ($\sim90\%$) have $\loiiicor<10^{40.5}~\ergs$.  After
all, LINERs are known to have low \oiiilam{} luminosities relative to
Seyfert 2 galaxies \citep[\eg,][]{heckman04a}.

\subsubsection{Galaxies with $\loiiicor>10^{40.5}~\ergs$ only}\label{sec:oiiionly}

\begin{figure}
\epsscale{1.0}
\plotone{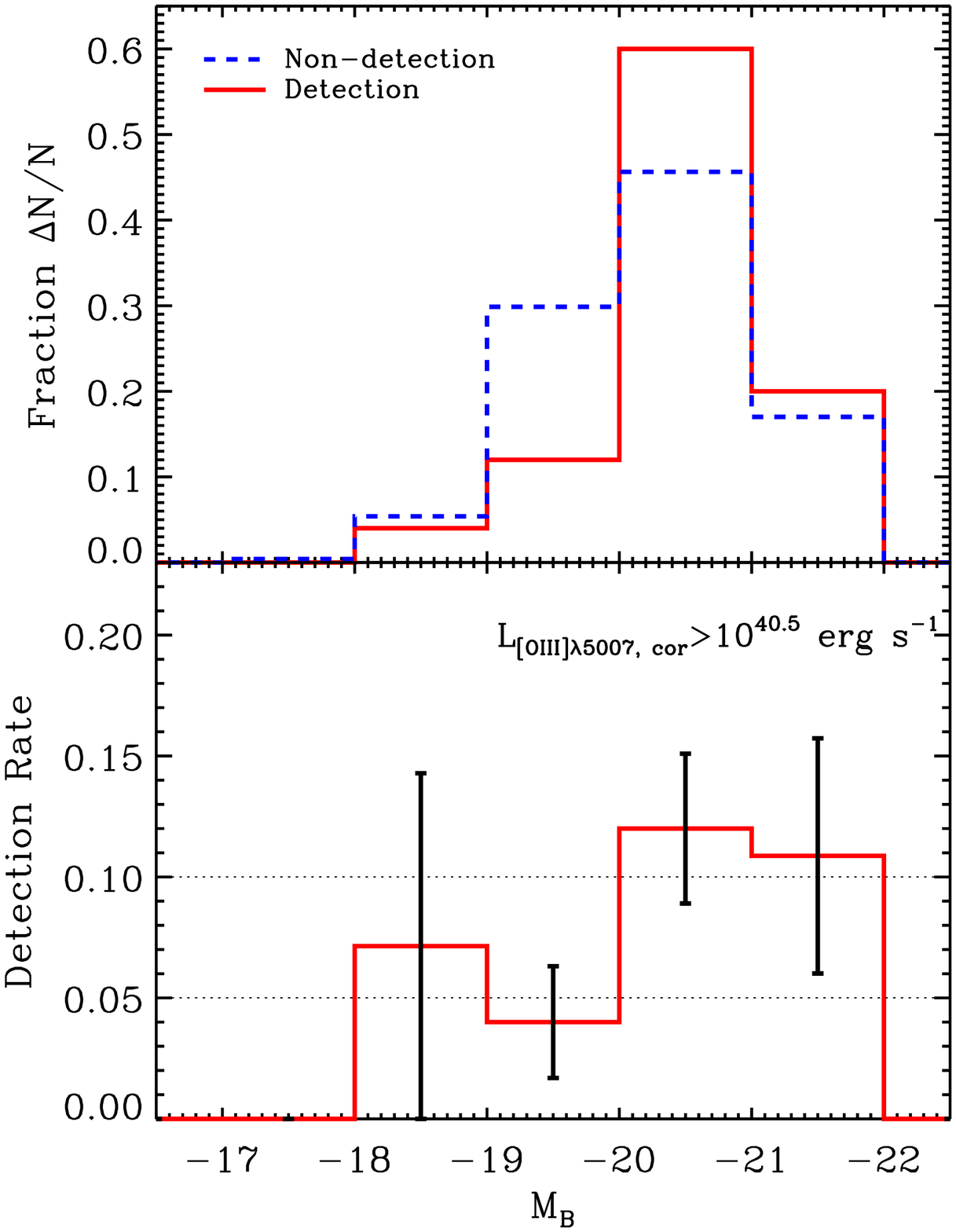}
\caption{
The same as the lower two panels in Figure \ref{fig:mbeff}, 
but for galaxies with $\loiiicor>10^{40.5}~\ergs$ only.
Note we double the binsize because of a smaller sample size. 
Among the $33$ detections, $25$ have $\loiiicor>10^{40.5}~\ergs$;
and among $1030$ non-detections, $242$ have $\loiiicor>10^{40.5}~\ergs$.
The overall detection rate ($\sim9\%$) is higher than that ($\sim3\%$) 
without pre-selection in Figure \ref{fig:mbeff}. 
The detection rate is also higher at higher luminosity.
}
\label{fig:mbeffsy2}
\end{figure}

\begin{figure}
\epsscale{1.0}
\plotone{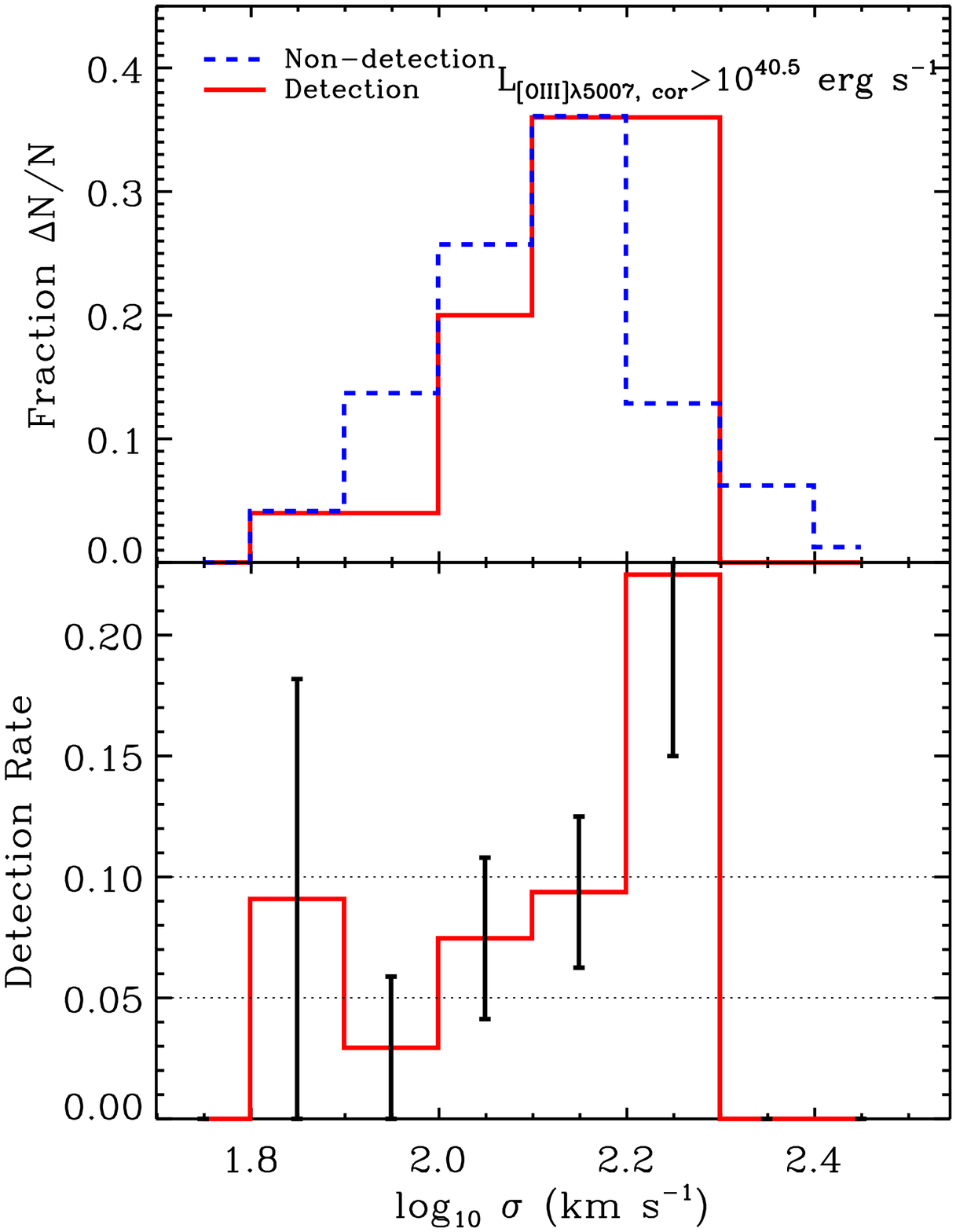}
\caption{
The same as the lower two panels in Figure \ref{fig:sigmaeff}, 
but for galaxies with $\loiiicor>10^{40.5}~\ergs$ only.
The overall detection rate is higher than without pre-selection
in Figure \ref{fig:sigmaeff}. 
The detection rate is higher at larger $\sigma$.
}
\label{fig:sigmaeffsy2}
\end{figure}

\begin{figure}
\epsscale{1.0}
\plotone{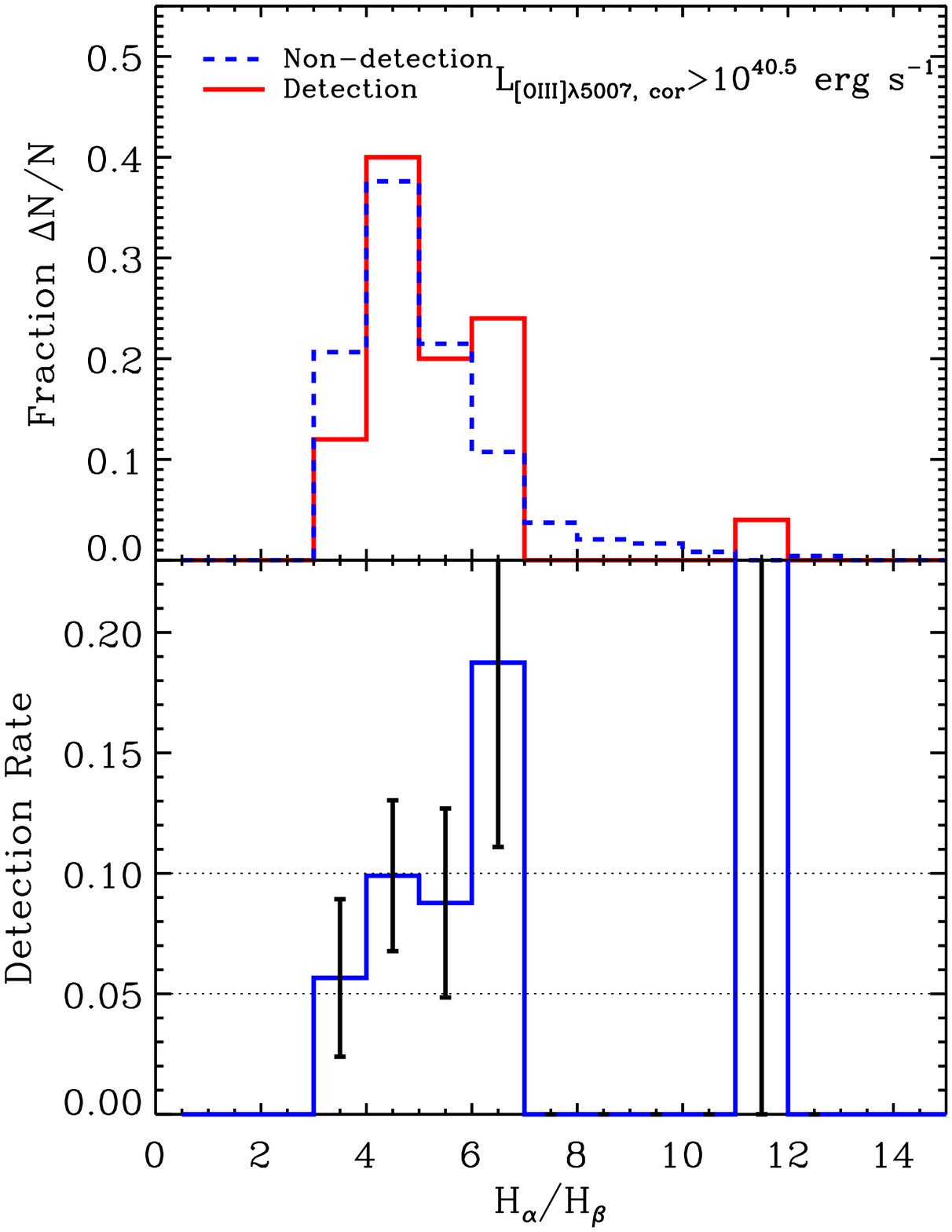}
\caption{
The same as the lower two panels in Figure \ref{fig:hahbeff}, 
but for galaxies with $\loiiicor>10^{40.5}~\ergs$ only.
The overall detection rate is higher than without pre-selection
in Figure \ref{fig:hahbeff}. 
The detection rate  is higher at higher $\Ha/\Hb$.
}
\label{fig:hahbeff2}
\end{figure}

We have shown that maser detection rate depends more 
strongly on $\loiiinocor$ than on $M_B$ and $\sigma$, as in 
Figures \ref{fig:mbeff}, \ref{fig:sigmaeff}, and \ref{fig:oiiieff}. 
However, these three quantities are themselves correlated in the
galaxy population. 
We here investigate further whether there is a residual dependence
of detection efficiency on $M_B$ and $\sigma$ even for galaxies with
strong \oiiilam{} emission.

Among the $33$ detections, only eight have $\loiiicor<10^{40.5}~\ergs$.
Meanwhile, $789$ out of $1030$ non-detections have
$\loiiicor<10^{40.5}~\ergs$.  We therefore pre-select galaxies with
$\loiiicor>10^{40.5}~\ergs$ ($25/33$ detections and $242/1030$
non-detections) and show the detection rate as a function of $M_B$ and
$\sigma$ for this pre-selected sample in Figure \ref{fig:mbeffsy2} and
\ref{fig:sigmaeffsy2}. We also show the detection rate as a function of 
Balmer decrement $\Ha/\Hb$ in Figure \ref{fig:hahbeff2}.
Compared to the results
for the whole sample (Figure \ref{fig:mbeff}, \ref{fig:sigmaeff}, and \ref{fig:hahbeff}),
there are still detections over the whole range of $M_B$, $\sigma$, and $\Ha/\Hb$,
and the detection rate is still higher at brighter $M_B$, 
larger $\sigma$, and higher $\Ha/\Hb$. 
The overall detection rate, however, is $\sim9\%$
compared to $\sim3\%$ without pre-selection.  
We therefore conclude that 
among $M_B$, $\sigma$, \oiiilam{}, and $\Ha/\Hb$, 
\oiiilam{} is the dominant factor when determining maser detection efficiency.

\section{Discussion}\label{sec:relations}

\subsection{Isotropic luminosity of masers}\label{sec:liso}


Above, we found that water maser detection rate increases with
\oiiilam~luminosity ($\loiiinocor$), velocity dispersion ($\sigma$), and
optical luminosity ($M_B$). We speculate that these correlations may
be a consequence of an underlying correlation of these parameters
with water maser luminosity. The true luminosity of masers is
difficult to measure because the maser emission is likely to
be beamed \citep[\eg, ][]{elitzur92a} 
and estimates of the beaming angle require a detailed model of the maser, which can only
be inferred in the cases with well-understood geometries from
VLBI observations \citep[\eg, ][]{miyoshi95a}. 
In place of true luminosity, we adopt apparent luminosity, which is based 
on the premise of ``isotropic'' emission of radiation. The isotropic luminosity 
can be computed readily from the flux densities observed in spectra.
On the other hand, the flux density can be variable on time scales of months 
\citep[\eg, ][]{bragg00a, braatz03a, herrnstein05a, castangia08a}, 
which introduces another uncertainty.
Nonetheless, analysis using the stand-in of isotropic luminosity provides
an opportunity to investigate whether the speculated correlations exist.
From the literature, we have collected isotropic luminosities for $66$ masers.

Maser surveys are flux-limited and the lower maser detection rate in galaxies 
with lower $\loiii$, smaller $\sigma$, and fainter $M_B$ could be because most
masers in these galaxies are too faint to be detected.
We investigate the flux limit in Figure \ref{fig:lisoz}, where 
we plot isotropic luminosities as a function of redshift.
Although these detections are from a variety of surveys that have different
sensitivities, they appear to be consistent with $0.1$ Jy $\kms$ as
the effective limit.  This limit is consistent with a plausible detection 
threshold of $10$ mJy ($3\sigma$) in a  $1~\kms$ channel and blends
of Doppler components on the order of $10~\kms$, as seen in the
spectra \citep[\eg, with GBT, ][]{braatz04a}.

\begin{figure}
\epsscale{1.1}
\plotone{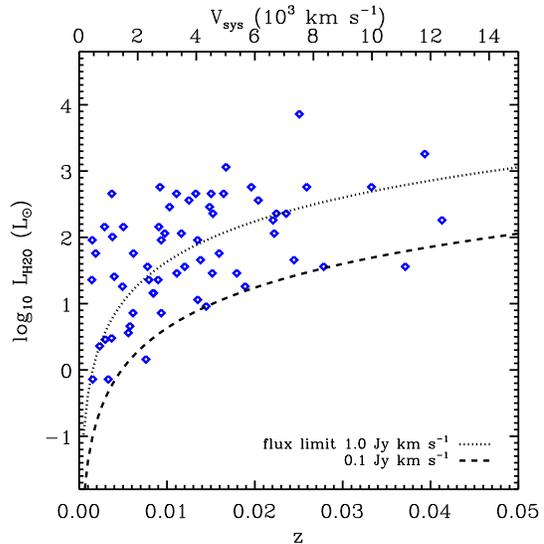}
\caption{
Isotropic maser luminosity as a function of redshift for $66$ extragalactic 
nuclear masers.  The dotted line shows the flux limit $1.0$ Jy $\kms$,
and the dashed line shows the flux limit $0.1$ Jy $\kms$.
}
\label{fig:lisoz}
\end{figure}

To investigate the correlations between maser luminosity and
optical properties of the host galaxies, 
we complement the sample with \mb, $\sigma$, \loiii, and $\Ha/\Hb$ 
(thus \loiiicor) from the literature. 
In addition to those with SDSS photometry or spectroscopy, we have compiled 
\loiiicor~($\Ha/\Hb$), \loiii, $\sigma$, and \mb~for $36$, $40$, $43$, 
and $54$ masers with measured isotropic luminosities. 
We present these data in Table \ref{table:data}.
The combined sample is inhomogeneous in terms of selection, technique, 
and instrument parameters, and this (in addition to the use of isotropic
luminosity as a stand-in -- noted earlier) may be expected to increase the
scatter in any correlations.

\begin{figure*}
\epsscale{1.1}
\plotone{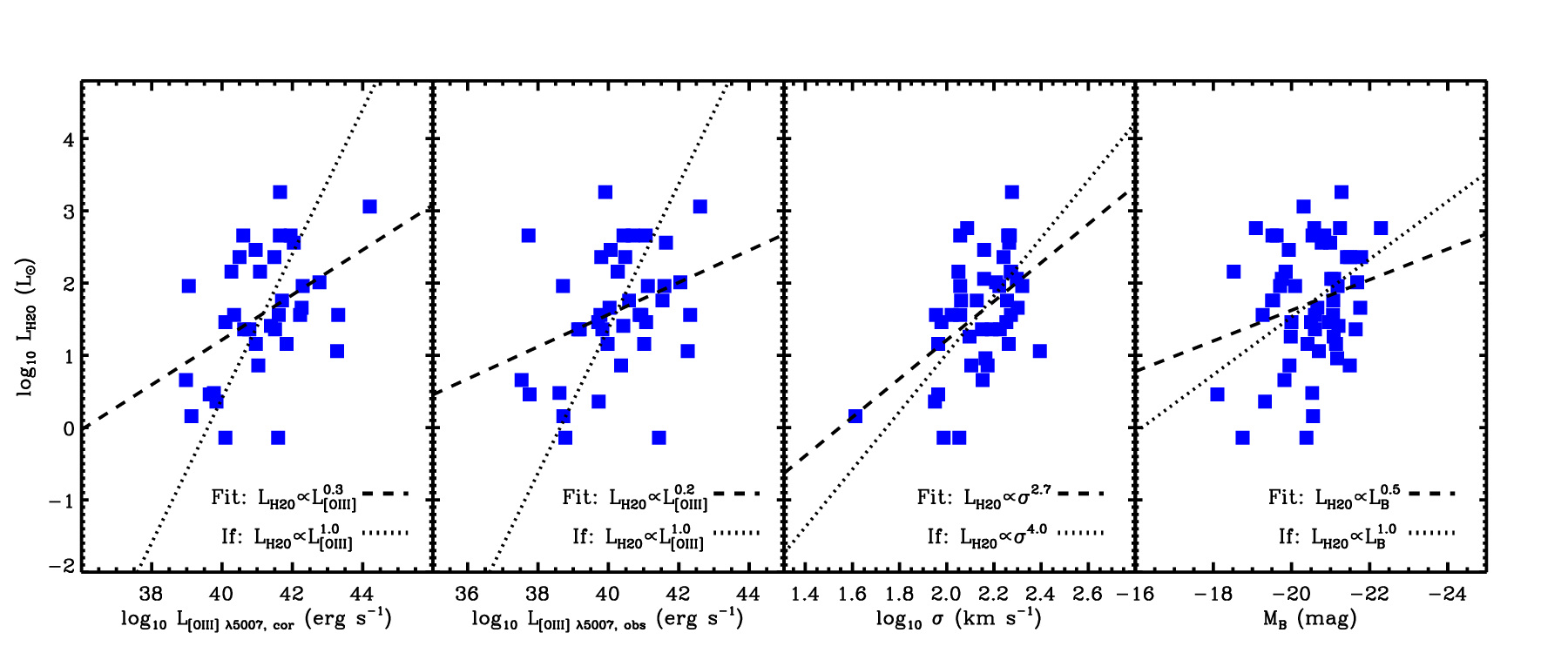}
\caption{
Relations between isotropic maser luminosity ($\lmaser$)
and \loiiicor, \loiii, $\sigma$, and 
\mb~of the host galaxies. The dashed lines are the linear least-squares fits 
assuming a uniform error of $0.5$ dex in $\log_{10} \lmaser$.
The dotted lines show the linear least-squares fits with the slopes fixed
assuming that $\lmaser \propto \lagn \propto \mbh$,
and $\mbh \propto L_B \propto \sigma^4$, and $\lagn \propto \loiiinocor$. 
}
\label{fig:liso}
\end{figure*}

\begin{figure}
\epsscale{1.0}
\plotone{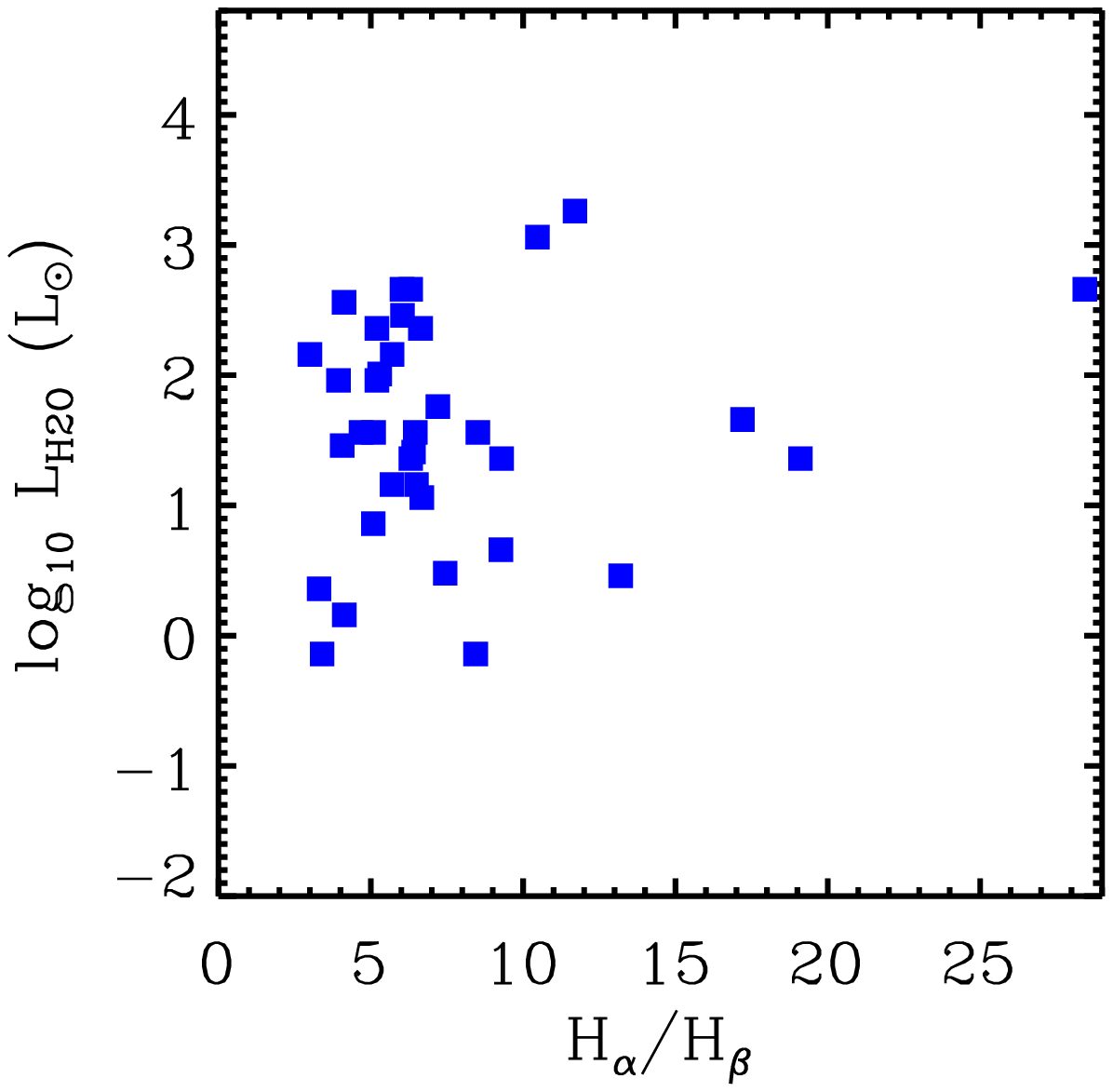}
\caption{
Relation between isotropic maser luminosity ($\lmaser$)
and $\Ha/\Hb$ of the host galaxies. 
}
\label{fig:liso2}
\end{figure}

In Figure \ref{fig:liso}, we plot $\log_{10} \loiiicor$, $\log_{10} \loiii$, $\log_{10} \sigma$, and \mb{} 
against the isotropic luminosity $\log_{10} \lmaser$.  
Although the scatter in all relations is relatively large, the variables are 
clearly correlated.  Assuming a uniform error of $0.5$ dex for $\log_{10} \lmaser$, 
we perform least-squares fits with the following linear relations: 
$\log_{10} \lmaser = a + b~(x-x_0)$,
where we choose $x_0$ to be the medians, $41.39$ (dex), $40.43$ (dex), $2.16$ (dex), 
and $-20.58$ (mag) for $x=\log_{10} \loiiicor$, $\log_{10} \loiii$, $\log_{10} \sigma$, and $\mb$, 
respectively, to minimize the correlation between $a$ and $b$.
We show the results with the dashed lines.
The intercepts and slopes ($a$, $b$) for 
$\log_{10} \loiiicor$, $\log_{10} \loiii$,  $\log_{10} \sigma$, and $M_B$ 
are $(1.65\pm0.13, 0.31\pm0.10)$, $(1.66\pm0.12,0.22\pm0.11)$, $(1.64\pm0.11,2.67\pm0.64)$, and 
$(1.75\pm0.11,-0.21\pm0.13)$, respectively.
We can draw a robust conclusion 
that masers are stronger
in hosts with higher \oiiilam~luminosity, larger $\sigma$, and higher 
optical luminosity, as implied from the detection efficiency results.

We would like to emphasize that the best-fit slopes and 
intercepts here should be taken with caution. 
First, as mentioned above, the isotropic luminosity is a poor indicator 
of maser strength and the sample of optical properties is not homogeneous.
Second, stronger masers are more easily detected, and at the faint end there 
should be a detection bias favoring stronger masers scattered above the real 
correlations; the real slopes therefore should be steeper than the best-fit ones.
Finally, the sample size is small and a few outliers could significantly affect
the fitting.  However, the conclusion is robust that maser strength increases
with \oiiilam~luminosity, $\sigma$, and optical luminosity.

In Figure \ref{fig:hahbeff}, we showed that maser detection rate is higher
at higher $\Ha/\Hb$. As for the other parameters, this could result from an underlying 
relation between maser luminosity and extinction.
We investigate such a relation in Figure \ref{fig:liso2}.
A linear least-squares fit gives 
($a$, $b$) = ($1.54\pm0.15$, $0.03\pm0.03$) for $x_0=6.31$.
We do not find a convincing relation (with only $\sim 1\sigma$) between 
maser luminosity and Balmer decrement, indicating that 
there might not be a direct linear relation between $\log_{10} \lmaser{}$ and 
Balmer decrement.
We discuss this result more in Section \ref{sec:geometry}.

\subsection{Correlations of maser emission with central black hole mass and AGN activity?}\label{sec:bhagn}

It has long been proposed that water maser emission is closely related 
with the central black hole.
Assuming that a thin viscous accretion disk is obliquely illuminated 
by a central X-ray source, \citet{neufeld95a} find that the 
critical outer radius \Rcr~at which the disk becomes atomic 
and the maser emission ceases follows 
$\Rcr \propto \lxray^{-0.43} \dot{m}^{0.81} \mbh^{0.62}$,
where \lxray~is the $2-10$ keV X-ray luminosity of the host galaxy,
$\dot{m}$ is the mass accretion rate, and
$\mbh$ is the central black hole mass.
If we assume that the X-ray luminosity is proportional to AGN 
bolometric luminosity ($L_{\rm AGN})$, and that $L_{\rm AGN}$ is proportional to 
the accretion rate \citep[\eg,][]{frank02a}, we consequently reach 
$\Rcr \propto \lagn^{0.38} \mbh^{0.62}$.
Considering that in the geometrical maser models \citep[\eg][]{miyoshi95a}, 
maser spots do not cover the whole nuclear disk but rather lie on several 
radial arms, we assume that total maser luminosity $\lmaser \propto \Rcr$
(but see \citealt{kondratko06a} who assume $\lmaser \propto \Rcr^2$).
We then expect $\lmaser \propto \lagn^{0.38} \mbh^{0.62}$.
If we further assume that AGNs radiate with a roughly fixed Eddington ratio
$\eta$ as a function of black hole mass, so that
$\lagn = \eta L_{\rm Edd} \propto \mbh$, we 
eventually reach $\lmaser \propto \lagn \propto \mbh$.

It is well-known that there 
exists a tight correlation between the central black hole mass ($\mbh$) 
and the velocity dispersion of the black hole host
\citep[][]{ferrarese00a, gebhardt00a, tremaine02a, gultekin09a}: 
$\mbh \propto \sigma^{4}$.
The optical luminosity also correlates with the black hole mass but with
a larger scatter \citep{magorrian98a, gultekin09a}: 
$\mbh \propto L (\rm Optical)$.
For AGN strength, the \oiiilam{} luminosity is a reasonably reliable 
indicator \citep[\eg, ][]{heckman05a}.
In Figure \ref{fig:liso},
assuming $\lmaser \propto \mbh \propto L_B \propto \sigma^4$ and 
$\lmaser \propto \lagn \propto \loiiinocor$,
we fit the intercepts (at $0$) and obtain 
$-39.6\pm0.2$, $-38.6\pm0.2$, $-7.0\pm0.1$, and $-6.5\pm0.1$, 
for $\log_{10} \loiiicor$, $\log_{10} \loiii$, $\log_{10} \sigma$, and $M_B$, respectively.
We show these fits with dotted lines.
If one takes our best-fit correlations in Section $3.1$ at face value, it implies
a weaker dependence of maser luminosity on black hole mass and 
AGN activity than the simplified theory.

The analysis above is simplified. For example, the Eddington ratio $\eta$
is not necessarily independent of black hole mass for masers, even if 
it is so for AGNs. If masers are powered by AGN luminosity,
then a high enough Eddington ratio is required to pump up a strong 
maser. This hypothesis can explain what we found in Section \ref{sec:oiiionly}, that
\oiiilam{} luminosity is more important a factor than $M_B$ and $\sigma$ 
when determining detection efficiency.
It is also interesting that the slopes above are all steeper than the 
best-fit ones. As we discussed in Section \ref{sec:liso}, the best-fit slopes
could be flatter than real ones due to the detection bias.
Considering the large uncertainties caused by the variability 
and the assumption of isotropy in maser luminosity and 
the small sample size,
we conclude that our fitting results are in reasonable agreement with
the simplified theory and that maser strength is indeed 
correlated with the central black hole mass and the AGN activity of
the host galaxies.

\begin{figure*}
\epsscale{1.0}
\plotone{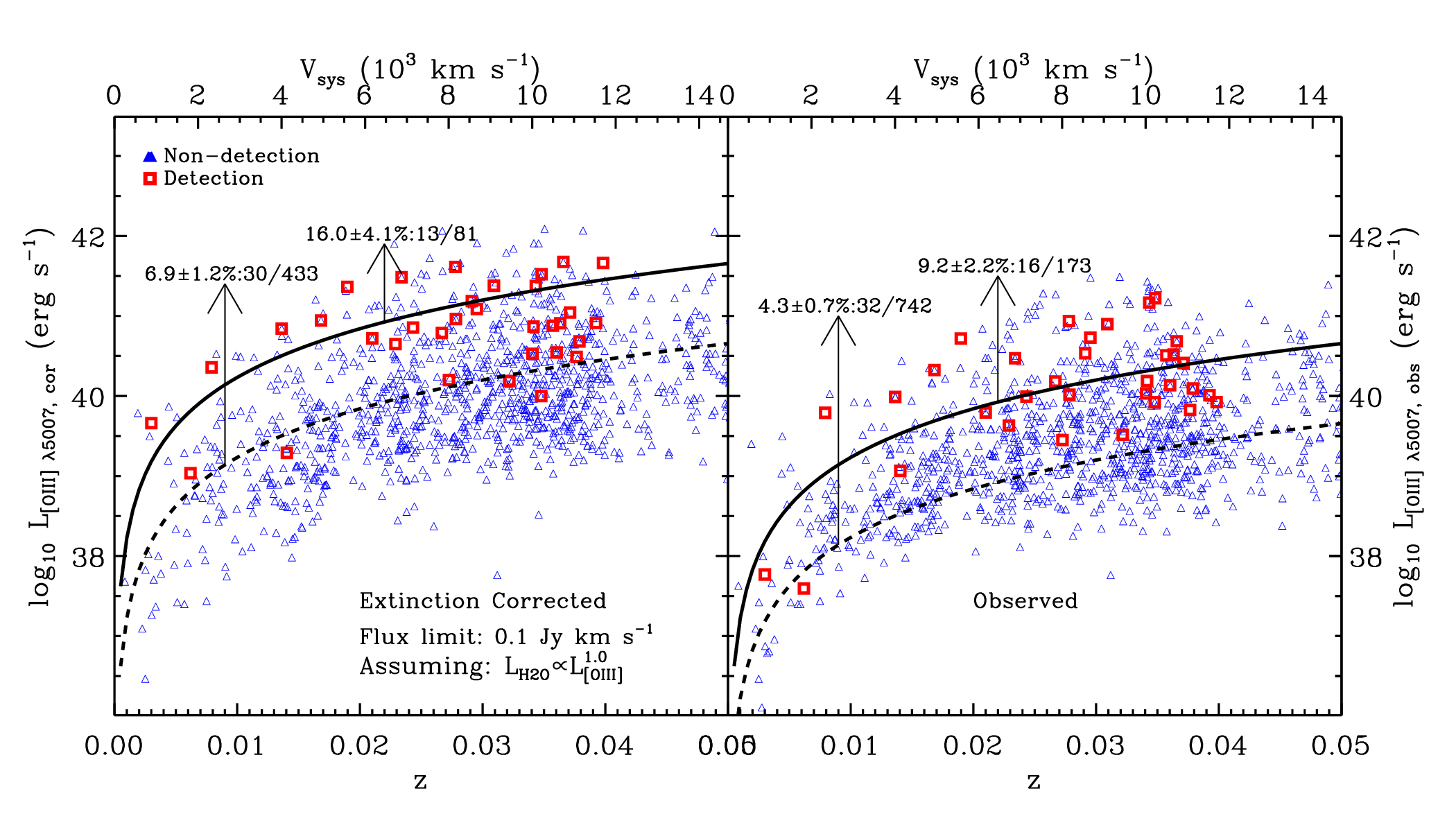}
\caption{
\oiiilam~luminosity as a function of redshift. We plot all detections
and non-detections in the SDSS spectroscopic sample. 
The solid lines show the assumption of maser flux limit $0.1$ Jy $\kms$ 
and $\lmaser \propto \loiiinocor$ (the dotted lines in Figure \ref{fig:liso}). 
The dashed lines are the flux limit shifted downward by $-1.0$ dex, to 
take into account the scatter in the \lmaser-\loiiinocor~relation.
The percentages at the end of arrows show the detection rates above
the corresponding lines and the two numbers show the number of 
detections and galaxies surveyed for maser emission (detections plus non-detections).
}
\label{fig:oiiiz}
\end{figure*}

Previous studies have also found correlations between the isotropic
luminosity and various properties of the host galaxies.
\citet{kondratko06a} find that $\lmaser \propto \sim \lxray^{2}$,
where \lxray~is the $2-10$ keV X-ray luminosity of the host galaxy.
\citet[see also \citealt{henkel05a, castangia08a, surcis09a}]{zhang10a} find
there appears to be a correlation between \lfir{} and \lmaser, where \lfir~is the total
far-infrared (FIR) luminosity of the host galaxy.  However, they do not claim to find a 
significant correlation between extinction-corrected \loiiicor{} and $\lmaser$, 
likely due to the smaller sample size in their study.

\subsection{Masers favor high-extinction systems: a geometrical effect?}\label{sec:geometry}

In Figure \ref{fig:hahbeff}, we showed that maser detection rate is 
higher at higher $\Ha/\Hb$, thus higher extinction. 
\citet{greenhill08a} and \citet{zhang10a} also find that there is a 
high incidence of Compton-thick systems among AGN masers. 
If masers are only located on the nuclear disk surrounding the central
black hole, then this preference of higher-extinction systems could be 
explained as a geometrical effect.

The specific intensity of maser emission scales with the cube of the pumping 
gain length ($l$) along the path \citep[\eg, ][]{strelnitskii84a, lo05a}: 
$\lmaser \propto l^3$.  If masers originate in gas clouds in the circumnuclear 
disk, we can only observe masers when the disk is inclined so that 
the gain length along the line-of-sight is long enough.
According to the proposed unified AGN models \citep[\eg, ][]{antonucci85a}, 
when the circumnuclear disk is inclined, the optical extinction due to the 
dusty torus and the X-ray obscuring column ($N_H$) are correspondingly higher.  
Naturally, masers are more often found to be in high-extinction systems.
Under this theory, we expect total isotropic maser luminosity on average 
to be higher in higher-extinction systems, since the average gain length 
should be longer. However, we did not find a convincing linear relation between 
$\log_{10} \lmaser$ and $\Ha/\Hb$ in Figure \ref{fig:liso2}.  This could be because 
$\Ha/\Hb$ is not a direct proxy of gain length, or the intrinsic scatter 
of the relation is too big to detect in the small sample we currently have.
Meanwhile, some masers, e.g., NGC $1052$ \citep[][]{claussen98a} and 
NGC $262$ \citep[Mrk $348$, ][]{peck03a}, are likely to be associated with 
jets but not disks. These systems could have greatly increased the scatter of the relation.
It is also possible that the inclination of the host galaxy can affect the
observed Balmer decrement, and thus increase the scatter.
On the other hand, if this theory is correct, then many extragalactic 
nuclear masers should be disk systems, but more than half of current maser 
detections lack evidence of association with disks (or jets).
To confirm this theory requires further investigations with larger samples and 
follow-up VLBI observations of maser detections.

\section{Suggestions on survey strategies}\label{sec:strategy}

Our results suggest that if we can improve the observational sensitivity, 
we should be able to detect more nuclear masers.
Given the current sensitivity, however, we can still improve the detection 
efficiency. Since there exist large optical spectroscopic surveys, such as 
SDSS, 2dFGRS, and 6dFGS,  we suggest that maser surveys primarily
interested in efficiency should select AGN targets from 
these surveys and rank them by extinction-corrected \oiiilam~flux.  
As an example, we illustrate this strategy in Figure \ref{fig:oiiiz}.

In Figure \ref{fig:oiiiz}, we plot $\log_{10} \loiiinocor$ against redshift for all 
the $33$ detections and $1030$ non-detections with SDSS spectroscopy.
Note in Figure \ref{fig:lisoz}, we showed that $0.1$ Jy $\kms$
represents the effective flux limit in maser surveys.  
To relate this limit to $\loiiinocor$, we use the dotted lines in 
Figure \ref{fig:liso}: $\log_{10} \lmaser = -39.6+\log_{10} \loiiicor$ and 
$\log_{10} \lmaser = -38.6+\log_{10} \loiii$,  
where $\lmaser$ is in $\Lsun$ and $\loiiinocor$ is in $\ergs$.
This conversion translates the limit $0.1$ Jy $\kms$ in maser flux 
into $7.6\times10^{-14}~\ergscm$ in extinction-corrected \oiiilam~flux, 
and $7.6\times10^{-15}~\ergscm$ in observed \oiiilam~flux. 
These limits are the solid lines in Figure \ref{fig:liso}. 
To take into account the scatter in the 
\lmaser-\loiiinocor~relation, we shift the limits downward by $-1.0$ dex to 
$7.6\times10^{-15}~\ergscm$ and $7.6\times10^{-16}~\ergscm$ and 
show them with the dashed lines. 
Among the $33$ detections, $30$ ($32$) have \loiiicor{} (\loiii) brighter than 
the dashed line.
For \loiiicor{} (\loiii),
the detection rates above and below the dashed line 
are $6.9\%\pm1.2\%$ ($4.3\%\pm0.8\%$) and $0.5\%\pm0.3\%$ ($0.3\%\pm0.3\%$);
the detection rates above and below the solid line are
$16.0\%\pm4.1\%$ ($9.2\%\pm2.2\%$) and $2.0\pm0.5\%$ ($1.9\%\pm0.5\%$).

Therefore, if we rank AGN targets by extinction-corrected \oiiilam~flux, 
we can significantly improve the detection efficiency, from an overall $\sim3\%$ 
to $\sim 16\%$ for the strongest \oiiilam~emitters.
Furthermore, given the current sensitivity, observing AGNs with 
extinction corrected \oiiilam~flux lower than the dashed line yields a very low 
detection rate of $\lesssim 1\%$. 
The contrast among these detection rates suggests that 
ranking source lists according to extinction-corrected \oiiilam{} 
is effective in maximizing detection efficiency.  
If extinction correction (i.e., $\Ha/\Hb$) is not available, we suggest 
that maser surveys rank AGN targets with the observed \oiiilam{} flux.
If the observed \oiiilam{} flux is not available either, we suggest that 
maser surveys rank AGN targets by velocity dispersion, or by optical luminosity.
This strategy should give a higher detection efficiency than a blind survey.

\acknowledgments

We are grateful to Jim Braatz for allowing us to use the compilation of 
the galaxy sample surveyed for water maser emission before publication.
We thank David W. Hogg and an anonymous referee for comments that helped 
improve the manuscript.
This research has made use of NASA’s Astrophysics Data System and of the 
NASA/IPAC Extragalactic Database (NED) which is operated by the Jet 
Propulsion Laboratory, California Institute of Technology, under contract 
with the National Aeronautics and Space Administration.  We also acknowledge 
the usage of the HyperLEDA database (http://leda.univ-lyon1.fr)
    
The authors acknowledge funding
support from NSF grant AST-0607701, NASA grants 06-FLEX06-0030,
NNX09AC85G and NNX09AC95G, and \emph{Spitzer} grant G05-AR-50443.
 
 Funding for the SDSS and SDSS-II has been provided by
 the Alfred P. Sloan Foundation,
 the Participating Institutions,
 the National Science Foundation,
 the U.S. Department of Energy,
 the National Aeronautics and Space Administration,
 the Japanese Monbukagakusho,
 the Max Planck Society,
 and the Higher Education Funding Council for England.
 The SDSS Web Site is http://www.sdss.org/.
 
%

\begin{thebibliography}{72}
\expandafter\ifx\csname natexlab\endcsname\relax\def\natexlab#1{#1}\fi

\bibitem[{{Abazajian} {et~al.}(2009){Abazajian}, {Adelman-McCarthy},
  {Ag{\"u}eros}, {Allam}, {Allende Prieto}, {An}, {Anderson}, {Anderson},
  {Annis}, {Bahcall}, \& et~al.}]{abazajian09a}
{Abazajian}, K.~N., {et~al.} 2009, \apjs, 182, 543

\bibitem[{{Adelman-McCarthy} {et~al.}(2006){Adelman-McCarthy}, {Ag{\"u}eros},
  {Allam}, {Anderson}, {Anderson}, {Annis}, {Bahcall}, {Baldry}, {Barentine},
  {Berlind}, {Bernardi}, {Blanton}, {Boroski}, {Brewington}, {Brinchmann},
  {Brinkmann}, {Brunner}, {Budav{\'a}ri}, {Carey}, {Carr}, {Castander},
  {Connolly}, {Csabai}, {Czarapata}, {Dalcanton}, {Doi}, {Dong}, {Eisenstein},
  {Evans}, {Fan}, {Finkbeiner}, {Friedman}, {Frieman}, {Fukugita}, {Gillespie},
  {Glazebrook}, {Gray}, {Grebel}, {Gunn}, {Gurbani}, {de Haas}, {Hall},
  {Harris}, {Harvanek}, {Hawley}, {Hayes}, {Hendry}, {Hennessy}, {Hindsley},
  {Hirata}, {Hogan}, {Hogg}, {Holmgren}, {Holtzman}, {Ichikawa}, {Ivezi{\'c}},
  {Jester}, {Johnston}, {Jorgensen}, {Juri{\'c}}, {Kent}, {Kleinman}, {Knapp},
  {Kniazev}, {Kron}, {Krzesinski}, {Kuropatkin}, {Lamb}, {Lampeitl}, {Lee},
  {Leger}, {Lin}, {Long}, {Loveday}, {Lupton}, {Margon},
  {Mart{\'{\i}}nez-Delgado}, {Mandelbaum}, {Matsubara}, {McGehee}, {McKay},
  {Meiksin}, {Munn}, {Nakajima}, {Nash}, {Neilsen}, {Newberg}, {Newman},
  {Nichol}, {Nicinski}, {Nieto-Santisteban}, {Nitta}, {O'Mullane}, {Okamura},
  {Owen}, {Padmanabhan}, {Pauls}, {Peoples}, {Pier}, {Pope}, {Pourbaix},
  {Quinn}, {Richards}, {Richmond}, {Rockosi}, {Schlegel}, {Schneider},
  {Schroeder}, {Scranton}, {Seljak}, {Sheldon}, {Shimasaku}, {Smith}, {Smol{\v
  c}i{\'c}}, {Snedden}, {Stoughton}, {Strauss}, {SubbaRao}, {Szalay},
  {Szapudi}, {Szkody}, {Tegmark}, {Thakar}, {Tucker}, {Uomoto}, {Vanden Berk},
  {Vandenberg}, {Vogeley}, {Voges}, {Vogt}, {Walkowicz}, {Weinberg}, {West},
  {White}, {Xu}, {Yanny}, {Yocum}, {York}, {Zehavi}, {Zibetti}, \&
  {Zucker}}]{adelman06a}
{Adelman-McCarthy}, J.~K., {et~al.} 2006, \apjs, 162, 38

\bibitem[{{Antonucci} \& {Miller}(1985)}]{antonucci85a}
{Antonucci}, R.~R.~J., \& {Miller}, J.~S. 1985, \apj, 297, 621

\bibitem[{{Baldwin} {et~al.}(1981){Baldwin}, {Phillips}, \&
  {Terlevich}}]{baldwin81a}
{Baldwin}, J.~A., {Phillips}, M.~M., \& {Terlevich}, R. 1981, \pasp, 93, 5

\bibitem[{{Balzano}(1983)}]{balzano83a}
{Balzano}, V.~A. 1983, \apj, 268, 602

\bibitem[{{Bassani} {et~al.}(1999){Bassani}, {Dadina}, {Maiolino}, {Salvati},
  {Risaliti}, {della Ceca}, {Matt}, \& {Zamorani}}]{bassani99a}
{Bassani}, L., {Dadina}, M., {Maiolino}, R., {Salvati}, M., {Risaliti}, G.,
  {della Ceca}, R., {Matt}, G., \& {Zamorani}, G. 1999, \apjs, 121, 473

\bibitem[{{Blanton} \& {Roweis}(2007)}]{blanton07a}
{Blanton}, M.~R., \& {Roweis}, S. 2007, \aj, 133, 734

\bibitem[{{Blanton} {et~al.}(2005){Blanton}, {Schlegel}, {Strauss},
  {Brinkmann}, {Finkbeiner}, {Fukugita}, {Gunn}, {Hogg}, {Ivezi{\'c}}, {Knapp},
  {Lupton}, {Munn}, {Schneider}, {Tegmark}, \& {Zehavi}}]{blanton05a}
{Blanton}, M.~R., {et~al.} 2005, \aj, 129, 2562

\bibitem[{{Braatz} \& {Gugliucci}(2008)}]{braatz08a}
{Braatz}, J.~A., \& {Gugliucci}, N.~E. 2008, \apj, 678, 96

\bibitem[{{Braatz} {et~al.}(2004){Braatz}, {Henkel}, {Greenhill}, {Moran}, \&
  {Wilson}}]{braatz04a}
{Braatz}, J.~A., {Henkel}, C., {Greenhill}, L.~J., {Moran}, J.~M., \& {Wilson},
  A.~S. 2004, \apjl, 617, L29

\bibitem[{{Braatz} {et~al.}(2010){Braatz}, {Reid}, {Humphreys}, {Henkel},
  {Condon}, \& {Lo}}]{braatz10a}
{Braatz}, J.~A., {Reid}, M.~J., {Humphreys}, E.~M.~L., {Henkel}, C., {Condon},
  J.~J., \& {Lo}, K.~Y. 2010, \apj, 718, 657

\bibitem[{{Braatz} {et~al.}(1996){Braatz}, {Wilson}, \& {Henkel}}]{braatz96a}
{Braatz}, J.~A., {Wilson}, A.~S., \& {Henkel}, C. 1996, \apjs, 106, 51

\bibitem[{{Braatz} {et~al.}(1997){Braatz}, {Wilson}, \& {Henkel}}]{braatz97a}
---. 1997, \apjs, 110, 321

\bibitem[{{Braatz} {et~al.}(2003){Braatz}, {Wilson}, {Henkel}, {Gough}, \&
  {Sinclair}}]{braatz03a}
{Braatz}, J.~A., {Wilson}, A.~S., {Henkel}, C., {Gough}, R., \& {Sinclair}, M.
  2003, \apjs, 146, 249

\bibitem[{{Bragg} {et~al.}(2000){Bragg}, {Greenhill}, {Moran}, \&
  {Henkel}}]{bragg00a}
{Bragg}, A.~E., {Greenhill}, L.~J., {Moran}, J.~M., \& {Henkel}, C. 2000, \apj,
  535, 73

\bibitem[{{Castangia} {et~al.}(2008){Castangia}, {Tarchi}, {Henkel}, \&
  {Menten}}]{castangia08a}
{Castangia}, P., {Tarchi}, A., {Henkel}, C., \& {Menten}, K.~M. 2008, \aap,
  479, 111

\bibitem[{{Claussen} {et~al.}(1998){Claussen}, {Diamond}, {Braatz}, {Wilson},
  \& {Henkel}}]{claussen98a}
{Claussen}, M.~J., {Diamond}, P.~J., {Braatz}, J.~A., {Wilson}, A.~S., \&
  {Henkel}, C. 1998, \apjl, 500, L129+

\bibitem[{{Colless} {et~al.}(2001){Colless}, {Dalton}, {Maddox}, {Sutherland},
  {Norberg}, {Cole}, {Bland-Hawthorn}, {Bridges}, {Cannon}, {Collins}, {Couch},
  {Cross}, {Deeley}, {De Propris}, {Driver}, {Efstathiou}, {Ellis}, {Frenk},
  {Glazebrook}, {Jackson}, {Lahav}, {Lewis}, {Lumsden}, {Madgwick}, {Peacock},
  {Peterson}, {Price}, {Seaborne}, \& {Taylor}}]{colless01a}
{Colless}, M., {et~al.} 2001, \mnras, 328, 1039

\bibitem[{{Corwin} {et~al.}(1994){Corwin}, {Buta}, \& {de
  Vaucouleurs}}]{corwin94a}
{Corwin}, Jr., H.~G., {Buta}, R.~J., \& {de Vaucouleurs}, G. 1994, \aj, 108,
  2128

\bibitem[{{Dahari} \& {De Robertis}(1988)}]{dahari88a}
{Dahari}, O., \& {De Robertis}, M.~M. 1988, \apjs, 67, 249

\bibitem[{{de Vaucouleurs} {et~al.}(1991){de Vaucouleurs}, {de Vaucouleurs},
  {Corwin}, {Buta}, {Paturel}, \& {Fouque}}]{devaucouleurs91a}
{de Vaucouleurs}, G., {de Vaucouleurs}, A., {Corwin}, Jr., H.~G., {Buta},
  R.~J., {Paturel}, G., \& {Fouque}, P. 1991, {Third Reference Catalogue of
  Bright Galaxies}

\bibitem[{{Diamond-Stanic} {et~al.}(2009){Diamond-Stanic}, {Rieke}, \&
  {Rigby}}]{diamond09a}
{Diamond-Stanic}, A.~M., {Rieke}, G.~H., \& {Rigby}, J.~R. 2009, \apj, 698, 623

\bibitem[{{Elitzur}(1992)}]{elitzur92a}
{Elitzur}, M., ed. 1992, Astrophysics and Space Science Library, Vol. 170,
  {Astronomical masers}

\bibitem[{{Ferrarese} \& {Merritt}(2000)}]{ferrarese00a}
{Ferrarese}, L., \& {Merritt}, D. 2000, \apjl, 539, L9

\bibitem[{{Frank, King, \& Raine}(2002)}]{frank02a}
{Frank, King, \& Raine}, ed. 2002, {Accretion Power in Astrophysics: Third
  Edition}

\bibitem[{{Gebhardt} {et~al.}(2000){Gebhardt}, {Bender}, {Bower}, {Dressler},
  {Faber}, {Filippenko}, {Green}, {Grillmair}, {Ho}, {Kormendy}, {Lauer},
  {Magorrian}, {Pinkney}, {Richstone}, \& {Tremaine}}]{gebhardt00a}
{Gebhardt}, K., {et~al.} 2000, \apjl, 539, L13

\bibitem[{{Gerssen} {et~al.}(2004){Gerssen}, {van der Marel}, {Axon}, {Mihos},
  {Hernquist}, \& {Barnes}}]{gerssen04a}
{Gerssen}, J., {van der Marel}, R.~P., {Axon}, D., {Mihos}, J.~C., {Hernquist},
  L., \& {Barnes}, J.~E. 2004, \aj, 127, 75

\bibitem[{{Greene} \& {Ho}(2006)}]{greene06a}
{Greene}, J.~E., \& {Ho}, L.~C. 2006, \apjl, 641, L21

\bibitem[{{Greenhill} {et~al.}(1997){Greenhill}, {Ellingsen}, {Norris},
  {Gough}, {Sinclair}, {Moran}, \& {Mushotzky}}]{greenhill97a}
{Greenhill}, L.~J., {Ellingsen}, S.~P., {Norris}, R.~P., {Gough}, R.~G.,
  {Sinclair}, M.~W., {Moran}, J.~M., \& {Mushotzky}, R. 1997, \apjl, 474, L103+

\bibitem[{{Greenhill} \& {Gwinn}(1997)}]{greenhill97b}
{Greenhill}, L.~J., \& {Gwinn}, C.~R. 1997, \apss, 248, 261

\bibitem[{{Greenhill} {et~al.}(2008){Greenhill}, {Tilak}, \&
  {Madejski}}]{greenhill08a}
{Greenhill}, L.~J., {Tilak}, A., \& {Madejski}, G. 2008, \apjl, 686, L13

\bibitem[{{Greenhill} {et~al.}(2003){Greenhill}, {Booth}, {Ellingsen},
  {Herrnstein}, {Jauncey}, {McCulloch}, {Moran}, {Norris}, {Reynolds}, \&
  {Tzioumis}}]{greenhill03a}
{Greenhill}, L.~J., {et~al.} 2003, \apj, 590, 162

\bibitem[{{Gu} {et~al.}(2006){Gu}, {Melnick}, {Cid Fernandes}, {Kunth},
  {Terlevich}, \& {Terlevich}}]{gu06a}
{Gu}, Q., {Melnick}, J., {Cid Fernandes}, R., {Kunth}, D., {Terlevich}, E., \&
  {Terlevich}, R. 2006, \mnras, 366, 480

\bibitem[{{G{\"u}ltekin} {et~al.}(2009){G{\"u}ltekin}, {Richstone}, {Gebhardt},
  {Lauer}, {Tremaine}, {Aller}, {Bender}, {Dressler}, {Faber}, {Filippenko},
  {Green}, {Ho}, {Kormendy}, {Magorrian}, {Pinkney}, \& {Siopis}}]{gultekin09a}
{G{\"u}ltekin}, K., {et~al.} 2009, \apj, 698, 198

\bibitem[{{Heckman} {et~al.}(2004){Heckman}, {Kauffmann}, {Brinchmann},
  {Charlot}, {Tremonti}, \& {White}}]{heckman04a}
{Heckman}, T.~M., {Kauffmann}, G., {Brinchmann}, J., {Charlot}, S., {Tremonti},
  C., \& {White}, S.~D.~M. 2004, \apj, 613, 109

\bibitem[{{Heckman} {et~al.}(2005){Heckman}, {Ptak}, {Hornschemeier}, \&
  {Kauffmann}}]{heckman05a}
{Heckman}, T.~M., {Ptak}, A., {Hornschemeier}, A., \& {Kauffmann}, G. 2005,
  \apj, 634, 161

\bibitem[{{Henkel} {et~al.}(2005){Henkel}, {Braatz}, {Tarchi}, {Peck}, {Nagar},
  {Greenhill}, {Wang}, \& {Hagiwara}}]{henkel05a}
{Henkel}, C., {Braatz}, J.~A., {Tarchi}, A., {Peck}, A.~B., {Nagar}, N.~M.,
  {Greenhill}, L.~J., {Wang}, M., \& {Hagiwara}, Y. 2005, \apss, 295, 107

\bibitem[{{Herrnstein} {et~al.}(2005){Herrnstein}, {Moran}, {Greenhill}, \&
  {Trotter}}]{herrnstein05a}
{Herrnstein}, J.~R., {Moran}, J.~M., {Greenhill}, L.~J., \& {Trotter}, A.~S.
  2005, \apj, 629, 719

\bibitem[{{Ho} {et~al.}(1997){Ho}, {Filippenko}, \& {Sargent}}]{ho97a}
{Ho}, L.~C., {Filippenko}, A.~V., \& {Sargent}, W.~L.~W. 1997, \apj, 487, 568

\bibitem[{{Ho} {et~al.}(2003){Ho}, {Filippenko}, \& {Sargent}}]{ho03a}
---. 2003, \apj, 583, 159

\bibitem[{{Ho} {et~al.}(2009){Ho}, {Greene}, {Filippenko}, \&
  {Sargent}}]{ho09a}
{Ho}, L.~C., {Greene}, J.~E., {Filippenko}, A.~V., \& {Sargent}, W.~L.~W. 2009,
  \apjs, 183, 1

\bibitem[{{Ishihara} {et~al.}(2001){Ishihara}, {Nakai}, {Iyomoto}, {Makishima},
  {Diamond}, \& {Hall}}]{ishihara01a}
{Ishihara}, Y., {Nakai}, N., {Iyomoto}, N., {Makishima}, K., {Diamond}, P., \&
  {Hall}, P. 2001, \pasj, 53, 215

\bibitem[{{Jones} {et~al.}(2004){Jones}, {Saunders}, {Colless}, {Read},
  {Parker}, {Watson}, {Campbell}, {Burkey}, {Mauch}, {Moore}, {Hartley},
  {Cass}, {James}, {Russell}, {Fiegert}, {Dawe}, {Huchra}, {Jarrett}, {Lahav},
  {Lucey}, {Mamon}, {Proust}, {Sadler}, \& {Wakamatsu}}]{jones04a}
{Jones}, D.~H., {et~al.} 2004, \mnras, 355, 747

\bibitem[{{Kauffmann} {et~al.}(2003){Kauffmann}, {Heckman}, {Tremonti},
  {Brinchmann}, {Charlot}, {White}, {Ridgway}, {Brinkmann}, {Fukugita}, {Hall},
  {Ivezi{\'c}}, {Richards}, \& {Schneider}}]{kauffmann03a}
{Kauffmann}, G., {et~al.} 2003, \mnras, 346, 1055

\bibitem[{{Kewley} {et~al.}(2001){Kewley}, {Dopita}, {Sutherland}, {Heisler},
  \& {Trevena}}]{kewley01a}
{Kewley}, L.~J., {Dopita}, M.~A., {Sutherland}, R.~S., {Heisler}, C.~A., \&
  {Trevena}, J. 2001, \apj, 556, 121

\bibitem[{{Kondratko} {et~al.}(2005){Kondratko}, {Greenhill}, \&
  {Moran}}]{kondratko05a}
{Kondratko}, P.~T., {Greenhill}, L.~J., \& {Moran}, J.~M. 2005, \apj, 618, 618

\bibitem[{{Kondratko} {et~al.}(2006{\natexlab{a}}){Kondratko}, {Greenhill}, \&
  {Moran}}]{kondratko06b}
---. 2006{\natexlab{a}}, \apj, 652, 136

\bibitem[{{Kondratko} {et~al.}(2006{\natexlab{b}}){Kondratko}, {Greenhill},
  {Moran}, {Lovell}, {Kuiper}, {Jauncey}, {Cameron}, {G{\'o}mez},
  {Garc{\'{\i}}a-Mir{\'o}}, {Moll}, {de Gregorio-Monsalvo}, \&
  {Jim{\'e}nez-Bail{\'o}n}}]{kondratko06a}
{Kondratko}, P.~T., {et~al.} 2006{\natexlab{b}}, \apj, 638, 100

\bibitem[{{Kuo} {et~al.}(2010){Kuo}, {Braatz}, {Condon}, {Impellizzeri}, {Lo},
  {Zaw}, {Schenker}, {Henkel}, {Reid}, \& {Greene}}]{kuo10a}
{Kuo}, C.~Y., {et~al.} 2010, ArXiv e-prints

\bibitem[{{Lo}(2005)}]{lo05a}
{Lo}, K.~Y. 2005, \araa, 43, 625

\bibitem[{{Madejski} {et~al.}(2006){Madejski}, {Done}, {{\.Z}ycki}, \&
  {Greenhill}}]{madejski06a}
{Madejski}, G., {Done}, C., {{\.Z}ycki}, P.~T., \& {Greenhill}, L. 2006, \apj,
  636, 75

\bibitem[{{Magorrian} {et~al.}(1998){Magorrian}, {Tremaine}, {Richstone},
  {Bender}, {Bower}, {Dressler}, {Faber}, {Gebhardt}, {Green}, {Grillmair},
  {Kormendy}, \& {Lauer}}]{magorrian98a}
{Magorrian}, J., {et~al.} 1998, \aj, 115, 2285

\bibitem[{{McElroy}(1995)}]{mcelroy95a}
{McElroy}, D.~B. 1995, \apjs, 100, 105

\bibitem[{{Miyoshi} {et~al.}(1995){Miyoshi}, {Moran}, {Herrnstein},
  {Greenhill}, {Nakai}, {Diamond}, \& {Inoue}}]{miyoshi95a}
{Miyoshi}, M., {Moran}, J., {Herrnstein}, J., {Greenhill}, L., {Nakai}, N.,
  {Diamond}, P., \& {Inoue}, M. 1995, \nat, 373, 127

\bibitem[{{Nelson} \& {Whittle}(1995)}]{nelson95a}
{Nelson}, C.~H., \& {Whittle}, M. 1995, \apjs, 99, 67

\bibitem[{{Neufeld} \& {Maloney}(1995)}]{neufeld95a}
{Neufeld}, D.~A., \& {Maloney}, P.~R. 1995, \apjl, 447, L17+

\bibitem[{{Oliva} {et~al.}(1995){Oliva}, {Origlia}, {Kotilainen}, \&
  {Moorwood}}]{oliva95a}
{Oliva}, E., {Origlia}, L., {Kotilainen}, J.~K., \& {Moorwood}, A.~F.~M. 1995,
  \aap, 301, 55

\bibitem[{{Oliva} {et~al.}(1999){Oliva}, {Origlia}, {Maiolino}, \&
  {Moorwood}}]{oliva99a}
{Oliva}, E., {Origlia}, L., {Maiolino}, R., \& {Moorwood}, A.~F.~M. 1999, \aap,
  350, 9

\bibitem[{{Peck} {et~al.}(2003){Peck}, {Henkel}, {Ulvestad}, {Brunthaler},
  {Falcke}, {Elitzur}, {Menten}, \& {Gallimore}}]{peck03a}
{Peck}, A.~B., {Henkel}, C., {Ulvestad}, J.~S., {Brunthaler}, A., {Falcke}, H.,
  {Elitzur}, M., {Menten}, K.~M., \& {Gallimore}, J.~F. 2003, \apj, 590, 149

\bibitem[{{Polletta} {et~al.}(1996){Polletta}, {Bassani}, {Malaguti},
  {Palumbo}, \& {Caroli}}]{polletta96a}
{Polletta}, M., {Bassani}, L., {Malaguti}, G., {Palumbo}, G.~G.~C., \&
  {Caroli}, E. 1996, \apjs, 106, 399

\bibitem[{{Sarzi} {et~al.}(2010){Sarzi}, {Shields}, {Schawinski}, {Jeong},
  {Shapiro}, {Bacon}, {Bureau}, {Cappellari}, {Davies}, {de Zeeuw}, {Emsellem},
  {Falc{\'o}n-Barroso}, {Krajnovi{\'c}}, {Kuntschner}, {McDermid}, {Peletier},
  {van den Bosch}, {van de Ven}, \& {Yi}}]{sarzi10a}
{Sarzi}, M., {et~al.} 2010, \mnras, 402, 2187

\bibitem[{{Schlegel} {et~al.}(1998){Schlegel}, {Finkbeiner}, \&
  {Davis}}]{schlegel98a}
{Schlegel}, D.~J., {Finkbeiner}, D.~P., \& {Davis}, M. 1998, \apj, 500, 525

\bibitem[{{Strelnitskii}(1984)}]{strelnitskii84a}
{Strelnitskii}, V.~S. 1984, \mnras, 207, 339

\bibitem[{{Surcis} {et~al.}(2009){Surcis}, {Tarchi}, {Henkel}, {Ott}, {Lovell},
  \& {Castangia}}]{surcis09a}
{Surcis}, G., {Tarchi}, A., {Henkel}, C., {Ott}, J., {Lovell}, J., \&
  {Castangia}, P. 2009, \aap, 502, 529

\bibitem[{{Tremaine} {et~al.}(2002){Tremaine}, {Gebhardt}, {Bender}, {Bower},
  {Dressler}, {Faber}, {Filippenko}, {Green}, {Grillmair}, {Ho}, {Kormendy},
  {Lauer}, {Magorrian}, {Pinkney}, \& {Richstone}}]{tremaine02a}
{Tremaine}, S., {et~al.} 2002, \apj, 574, 740

\bibitem[{{Tremonti} {et~al.}(2004){Tremonti}, {Heckman}, {Kauffmann},
  {Brinchmann}, {Charlot}, {White}, {Seibert}, {Peng}, {Schlegel}, {Uomoto},
  {Fukugita}, \& {Brinkmann}}]{tremonti04a}
{Tremonti}, C.~A., {et~al.} 2004, \apj, 613, 898

\bibitem[{{Veilleux} \& {Osterbrock}(1987)}]{veilleux87a}
{Veilleux}, S., \& {Osterbrock}, D.~E. 1987, \apjs, 63, 295

\bibitem[{{Wegner} {et~al.}(2003){Wegner}, {Bernardi}, {Willmer}, {da Costa},
  {Alonso}, {Pellegrini}, {Maia}, {Chaves}, \& {Rit{\'e}}}]{wegner03a}
{Wegner}, G., {et~al.} 2003, \aj, 126, 2268

\bibitem[{{Whittle}(1992)}]{whittle92a}
{Whittle}, M. 1992, \apjs, 79, 49

\bibitem[{{York} {et~al.}(2000){York}, {Adelman}, {Anderson}, {Anderson},
  {Annis}, {Bahcall}, {Bakken}, {Barkhouser}, {Bastian}, {Berman}, {Boroski},
  {Bracker}, {Briegel}, {Briggs}, {Brinkmann}, {Brunner}, {Burles}, {Carey},
  {Carr}, {Castander}, {Chen}, {Colestock}, {Connolly}, {Crocker}, {Csabai},
  {Czarapata}, {Davis}, {Doi}, {Dombeck}, {Eisenstein}, {Ellman}, {Elms},
  {Evans}, {Fan}, {Federwitz}, {Fiscelli}, {Friedman}, {Frieman}, {Fukugita},
  {Gillespie}, {Gunn}, {Gurbani}, {de Haas}, {Haldeman}, {Harris}, {Hayes},
  {Heckman}, {Hennessy}, {Hindsley}, {Holm}, {Holmgren}, {Huang}, {Hull},
  {Husby}, {Ichikawa}, {Ichikawa}, {Ivezi{\'c}}, {Kent}, {Kim}, {Kinney},
  {Klaene}, {Kleinman}, {Kleinman}, {Knapp}, {Korienek}, {Kron}, {Kunszt},
  {Lamb}, {Lee}, {Leger}, {Limmongkol}, {Lindenmeyer}, {Long}, {Loomis},
  {Loveday}, {Lucinio}, {Lupton}, {MacKinnon}, {Mannery}, {Mantsch}, {Margon},
  {McGehee}, {McKay}, {Meiksin}, {Merelli}, {Monet}, {Munn}, {Narayanan},
  {Nash}, {Neilsen}, {Neswold}, {Newberg}, {Nichol}, {Nicinski}, {Nonino},
  {Okada}, {Okamura}, {Ostriker}, {Owen}, {Pauls}, {Peoples}, {Peterson},
  {Petravick}, {Pier}, {Pope}, {Pordes}, {Prosapio}, {Rechenmacher}, {Quinn},
  {Richards}, {Richmond}, {Rivetta}, {Rockosi}, {Ruthmansdorfer}, {Sandford},
  {Schlegel}, {Schneider}, {Sekiguchi}, {Sergey}, {Shimasaku}, {Siegmund},
  {Smee}, {Smith}, {Snedden}, {Stone}, {Stoughton}, {Strauss}, {Stubbs},
  {SubbaRao}, {Szalay}, {Szapudi}, {Szokoly}, {Thakar}, {Tremonti}, {Tucker},
  {Uomoto}, {Vanden Berk}, {Vogeley}, {Waddell}, {Wang}, {Watanabe},
  {Weinberg}, {Yanny}, \& {Yasuda}}]{york00a}
{York}, D.~G., {et~al.} 2000, \aj, 120, 1579

\bibitem[{{Zhang} {et~al.}(2010){Zhang}, {Henkel}, {Guo}, {Wang}, \&
  {Fan}}]{zhang10a}
{Zhang}, J.~S., {Henkel}, C., {Guo}, Q., {Wang}, H.~G., \& {Fan}, J.~H. 2010,
  \apj, 708, 1528

\bibitem[{{Zhang} {et~al.}(2006){Zhang}, {Henkel}, {Kadler}, {Greenhill},
  {Nagar}, {Wilson}, \& {Braatz}}]{zhang06a}
{Zhang}, J.~S., {Henkel}, C., {Kadler}, M., {Greenhill}, L.~J., {Nagar}, N.,
  {Wilson}, A.~S., \& {Braatz}, J.~A. 2006, \aap, 450, 933

\end{thebibliography}

\newpage
\clearpage
\LongTables
\begin{deluxetable*}{lcccccccccc}
\tabletypesize{\scriptsize}
\tablecaption{Physical parameters of extragalactic water maser sources}
\tablehead{
\colhead{Source} & \colhead{V$_{\rm sys}$\tablenotemark{1}} &\colhead{L$_{\rm H2O}$\tablenotemark{2}}  & \colhead{Ref(L$_{\rm H2O}$)\tablenotemark{3}} 
& \colhead{F$_{\rm [OIII]}$\tablenotemark{4}} & \colhead{H$_{\rm \alpha}$/H$_{\rm \beta}$ \tablenotemark{5}} 
& \colhead{Ref(F$_{\rm [OIII]}$)\tablenotemark{6}} 
&  \colhead{$\sigma$\tablenotemark{7}} & \colhead{Ref($\sigma$)\tablenotemark{8}}
& \colhead{M$_{B}$\tablenotemark{9}} & \colhead{Ref(M$_{B}$)\tablenotemark{10}}}
\startdata   
        NGC 23 (MrK 545)                           &   4566   & 2.3 & BG08                 &     1200   & 5.2  &  DD88       &    ... &  ...       & -21.50 &      DeV91           \\
        NGC 17 (Mrk 938)$^\dagger$                 &   5881   & ... & ...                  &      ...   & ...  &  ...        &    ... &  ...       &    ... &      ...             \\
        2MASX J00114518-0054303                    &  14384   & ... & ...                  &      ...   & ...  &  ...        &    ... &  ...       & -19.58 &      SDSS            \\
        NGC 235A                                   &   6664   & 2.0 & Kondratko06          &      ...   & ...  &  ...        & 198    &  Wegner03  & -21.02 &      DeV91           \\
        NGC 262 (Mrk 348)                          &   4507   & 2.6 & Henkel05             &    22710   & 6.02 &  Bassani99  & 185    &  McElroy95 & -20.54 &      DeV91           \\
        NGC 291                                    &   5705   & ... & ...                  &     6431   & 4.97 &  SDSS       & 109    &  SDSS      & -20.14 &      SDSS            \\
        ESO 013-G012                               &   5047   & ... & ...                  &      ...   & ...  &  ...        &    ... &  ...       &    ... &      ...             \\
        NGC 449 (Mrk 1)$^\dagger$                  &   4780   & 1.7 & Henkel05             &    60000   & ...  &  Whittle92  & 115    &  NW95      & -19.53 &      DeV91           \\
        NGC 520                                    &   2281   & 0.1 & Castangia08          &      412.4 & 4.13 &  Ho97       & 41     &  Ho09      & -20.55 &      DeV91           \\
        2MASX J01260163-0417564                    &   5639   & ... & ...                  &      ...   & ...  &  ...        &    ... &  ...       &    ... &      ...             \\
        NGC 591 (MrK 1157)$^\dagger$               &   4547   & 1.4 & Henkel05             &    23000   & ...  &  Whittle92  &  95    &  NW95      & -20.50 &      DeV91           \\
        NGC 613                                    &   1481   & 1.2 & Kondratko06          &      ...   & ...  &  ...        & 125    &  McElroy95 & -21.08 &      DeV91           \\
                                                   &          & 1.2 &         Castangia08  &            & ...  &             &        &            &        &                      \\
        IC 0184                                    &   5382   & 1.4 & Kondratko06          &      ...   & ...  &  ...        &    ... &  ...       & -20.00 &      DeV91           \\
        2MASX J02140591-0016371                    &  11205   & ... & ...                  &      ...   & ...  &  ...        &    ... &  ...       & -18.59 &      SDSS            \\
        Mrk 1029                                   &   9076   & ... & ...                  &      ...   & ...  &  ...        &    ... &  ...       &    ... &      ...             \\
        NGC 1052                                   &   1510   & 2.1 & Henkel05             &    33068   & 2.35\tablenotemark{b} &  Ho97       & 187    &  Wegner03  & -19.85 &      SDSS            \\
        NGC 1068 (M 77)$^\dagger$                  &   1137   & 2.2 & Henkel05             &  3497963   & 5.29 &  Ho97       & 162    &  Ho09      & -21.68 &      DeV91           \\
                                                   &          & 1.7 &            KGM06     &            &      &             &        &            &        &                      \\
        NGC 1106                                   &   4337   & 0.9 & BG08                 &        ... & ...  &  ...        & 146    &  Wegner03  & -21.17 &      DeV91           \\
        2MASX J02532956-0014052                    &   8622   & ... & ...                  &     1757   & 5.0  &  SDSS       &  96    &  SDSS      & -18.18 &      SDSS            \\
        Mrk 1066                                   &   3605   & 1.5 & Henkel05             &    24000   & 8.51 &  Whittle92  & 105    &  NW95      & -20.60 &      DeV91           \\
        NGC 1194$^\dagger$                         &   4076   & ... & ...                  &     2340   & 5.85 &  SDSS       & 144    &  SDSS      & -20.06 &      SDSS            \\
        NGC 1320 (MrK 607)$^\dagger$               &   2663   & ... & ...                  &        ... & ...  &  ...        &    ... &  ...       &    ... &      ...             \\
        2MASX J03364614-0750236                    &  11719   & ... & ...                  &        ... & ...  &  ...        &    ... &  ...       &    ... &      ...             \\
        NGC 1386$^\dagger$                         &    868   & 2.1 & Henkel05             &    99244   & 5.7  &  Bassani99  & 187    &  McElroy95 & -18.52 &      DeV91           \\
        2MASX J03381036+0114178                    &  11926   & ... & ...                  &        ... & ...  &  ...        &    ... &  ...       &    ... &      ...             \\
        (IRAS03355+0104)                           &          &     &                      &            &      &             &        &            &        &                      \\
        4C +05.19                                  & 790000   & ... & ...                  &        ... & ...  &  ...        &    ... &  ...       &    ... &      ...             \\
        (2MASX J04143774+0534423)                  &          &     &                      &            &      &             &        &            &        &                      \\
        2MASX J04405494-0822221                    &   4527   & ... & ...                  &        ... & ...  &  ...        &    ... &  ...       &    ... &      ...             \\
        (IRAS F04385-0828)                         &          &     &                      &            &      &             &        &            &        &                      \\
        UGC 3193$^\dagger$                         &   4454   & 2.4 & BG08                 &        ... & ...  &  ...        &    ... &  ...       & -19.93 &      DeV91           \\
        NGC 1741 (Mrk 1089)                        &   4039   & ... & ...                  &        ... & ...  &  ...        &    ... &  ...       &    ... &      ...             \\
        CGCG 468-002                               &   5454   & ... & ...                  &        ... & ...  &  ...        &    ... &  ...       &    ... &      ...             \\
        UGC 3255                                   &   5669   & 1.2 & Henkel05             &        ... & ...  &  ...        &    ... &  ...       & -19.98 &      DeV91           \\
        UGCA 116                                   &    789   & ... & ...                  &        ... & ...  &  ...        &    ... &  ...       &    ... &      ...             \\
        Mrk 3                                      &   4050   & 1.0 & Henkel05             &   439100   & 6.67 &  Bassani99  & 248    &  McElroy95 & -20.70 &      DeV91           \\
        VII Zw 073                                 &  12391   & 2.2 & Kondratko06          &        ... & ...  &  ...        &    ... &  ...       &    ... &      ...             \\
        NGC 2273 (MrK 620)                         &   1840   & 0.8 & Zhang06              &    27618   & 5.08 &  Ho97       & 149    &  Ho09      & -19.95 &      DeV91           \\
        UGC 3789$^\dagger$                         &   3325   & 2.6 & BG08                 &        ... & ...  &  ...        &    ... &  ...       & -20.47 &      DeV91           \\
        NGC 2410                                   &   4681   & ... & ...                  &     3306   & 4.88 &  SDSS       & 166    &  SDSS      & -20.86 &      SDSS            \\
        Mrk 78                                     &  11137   & 1.5 & Henkel05             &    66000   & 6.46 &  Polletta96 & 114    &  McElroy95 &    ... &      ...             \\
        IC 0485$^\dagger$                          &   8338   & ... & ...                  &      585.8 & 6.29 &  SDSS       & 187    &  SDSS      & -20.24 &      SDSS            \\
        Mrk 1210 (UGC 04203, Phoenix)              &   4046   & 1.9 & Henkel05             &    95660   & 5.2  &  Bassani99  & 114    &  Gu06      & -19.71 &      DeV91           \\
        SDSS J0804+3607                            & 198000   & ... & ...                  &        ... & ...  &  ...        &    ... &  ...       &    ... &      ...             \\
        2MASX J08362280+3327383$^\dagger$          &  14810   & ... & ...                  &        ... & ...  &  ...        &    ... &  ...       & -20.24 &      SDSS            \\
        NGC 2639$^\dagger$                         &   3336   & 1.4 & Henkel05             &     1858   & 4.06 &  Ho97       & 179    &  Ho09      & -20.95 &      SDSS            \\
        NGC 2781                                   &   2053   & ... & ...                  &        ... & ...  &  ...        &    ... &  ...       &    ... &      ...             \\
        2MASX J09124641+2304273                    &  10861   & ... & ...                  &     1078   & 4.07 &  SDSS       &  74    &  SDSS      & -19.12 &      SDSS            \\
        NGC 2782                                   &   2543   & 1.1 & Henkel05             &     5944   & 6.5  &  Ho97       & 183    &  Ho09      & -20.41 &      SDSS            \\
        NGC 2824 (MrK 394)                         &   2760   & 2.7 & Henkel05             &        ... & ...  &  ...        & 122    &  McElroy95 & -19.09 &      SDSS            \\
        SBS 0927+493                               &  10167   & ... & ...                  &      570.9 & 5.11 &  SDSS       & 147    &  SDSS      & -20.70 &      SDSS            \\
        UGC 5101                                   &  11802   & 3.2 & Zhang06              &      226.9 & 11.7 &  SDSS       & 189    &  SDSS      & -21.28 &      SDSS            \\
        NGC 2960 (MrK 1419)$^\dagger$              &   4932   & 2.6 & Henkel05             &     4400   & ...  &  DD88       &    ... &  ...       & -20.84 &      SDSS            \\
        NGC 2979$^\dagger$                         &   2720   & 2.1 & Henkel05             &        ... & ...  &  ...        & 112    &  Gu06      &    ... &      ...             \\
        NGC 2989                                   &   4146   & 1.6 & BG08                 &        ... & ...  &  ...        &    ... &  ...       & -20.66 &      DeV91           \\
        NGC 3081                                   &   2391   & ... & ...                  &        ... & ...  &  ...        &    ... &  ...       &    ... &      ...             \\
        NGC 3079$^\dagger$                         &   1116   & 2.7 & Henkel05             &      176.2 & 28.4 &  Ho97       & 182    &  Ho09      & -19.62 &      DeV91           \\
                                                   &          & 2.5 & Kondratko05          &            &      &             &        &            &        &                      \\
        2MASX J10115058-1926436                    &   8065   & ... & ...                  &        ... & ...  &  ...        &    ... &  ...       &    ... &      ...             \\
        NGC 3160                                   &   6920   & ... & ...                  &      358.5 & 6.66 &  SDSS       & 155    &  SDSS      & -20.32 &      SDSS            \\
        IC 2560$^\dagger$                          &   2925   & 2.0 & Henkel05             &        ... & ...  &  ...        & 144    &  Gu06      & -21.09 &      DeV91           \\
        NGC 3256                                   &   2804   & 0.8 & Surcis09             &        ... & ...  &  ...        & 127    &  Oliva95   & -21.49 &      DeV91           \\
        UGC 5713$^\dagger$                         &   6312   & ... & ...                  &      622.0 & 6.2  &  SDSS       & 168    &  SDSS      & -20.24 &      SDSS            \\
        Mrk 34$^\dagger$                           &  15140   & 3.0 & Henkel05             &    67000   & 10.5 &  Polletta96 &    ... &  ...       &    ... &      ...             \\
        NGC 3393$^\dagger$                         &   3750   & 2.4 & Kondratko06          &   124344   & 4.12 &  Bassani99  & 184    &  McElroy95 & -20.99 &      DeV91           \\
                                                   &          & 2.6 &         Zhang06      &            &      &             &        &            &        &                      \\
        UGC 6093$^\dagger$                         &  10828   & ... & ...                  &      452.0 & 4.14 &  SDSS       & 160    &  SDSS      & -21.53 &      SDSS            \\
        2MASX J11093314+2837393                    &  11422   & ... & ...                  &      371.7 & 4.76 &  SDSS       & 149    &  SDSS      & -20.41 &      SDSS            \\
        NGC 3620                                   &   1680   & 0.5 & Surcis09             &        ... & ...  &  ...        &    ... &  ...       &    ... &      ...             \\
        CGCG 185-028                               &  10455   & ... & ...                  &      298.7 & 3.19 &  SDSS       & 204    &  SDSS      & -21.14 &      SDSS            \\
        NGC 3614                                   &   2333   & ... & ...                  &        ... & ...  &  ...        &    ... &  ...       & -19.68 &      SDSS            \\
        Arp 299 (NGC 3690 $\&$ IC 694)\tablenotemark{a} &   3088   & 2.4 & Henkel05        &     4917   & 6.04 &  Ho97       & 144    &  Ho09      &    ... &      ...             \\
        NGC 3735$^\dagger$                         &   2696   & 1.3 & Henkel05             &     3741   & 6.31 &  Ho97       & 141    &  Ho09      & -20.60 &      DeV91           \\
        CGCG 068-013                               &  10660   & ... & ...                  &     1092   & 4.0  &  SDSS       & 111    &  SDSS      & -21.28 &      SDSS            \\
        NGC 3783                                   &   2917   & ... & ...                  &        ... & ...  &  ...        &    ... &  ...       &    ... &      ...             \\
        CGCG 268-089                               &   7924   & ... & ...                  &      930.5 & 4.83 &  SDSS       & 128    &  SDSS      & -20.22 &      SDSS            \\
        (MCG +09-19-205)                           &          &     &                      &            &      &             &        &            &        &                      \\
        2MASX J12020465+3519173$^\dagger$          &  10201   & ... & ...                  &      401.0 & 4.43 &  SDSS       & 105    &  SDSS      & -19.94 &      SDSS            \\
        UGC 7016$^\dagger$                         &   7271   & ... & ...                  &      728.6 & 4.89 &  SDSS       & 177    &  SDSS      & -20.83 &      SDSS            \\
        NGC 4051$^\dagger$                         &    700   & 0.3 & Henkel05             &    44009   & 3.3  &  Ho97       &  89    &  GH06      & -19.32 &      DeV91           \\
        NGC 4151                                   &    995   &-0.2 & Henkel05             &  1125847   & 3.4  &  Ho97       &  97    &  GH06      & -18.75 &      SDSS            \\
        NGC 4253 (MrK 766)                         &   3876   & ... & ...                  &    262.0   & 3.58 &  SDSS       &  85    &  SDSS      & -20.34 &      SDSS            \\
        NGC 4258 (M 106)$^\dagger$                 &    448   & 1.9 & Henkel05             &    10430   & 3.94 &  Ho97       & 167    &  McElroy95 & -20.09 &      DeV91           \\
        NGC 4293                                   &    893   & 0.4 & Kondratko06          &      295.3 & 13.2 &  SDSS       & 92     &  SDSS      & -18.10 &      SDSS            \\
        NGC 4388$^\dagger$                         &   2524   & 1.1 & Henkel05             &    66226   & 5.69 &  Ho97       & 92     &  Ho09      & -21.14 &      SDSS            \\
        NGC 4527                                   &   1736   & 0.6 & BG08                 &      46.30 & 9.27 &  SDSS       & 142    &  SDSS      & -19.82 &      SDSS            \\
        NGC 4633 (IC 3688)                         &    291   & ... & ...                  &        ... & ...  &  ...        &    ... &  ...       &    ... &      ...             \\
        ESO 269-G012$^\dagger$                     &   5014   & 3.0 & Henkel05             &        ... & ...  &  ...        &    ... &  ...       & -20.31 &      DeV91           \\
        NGC 4922N                                  &   7071   & 2.3 & Henkel05             &     2379   & 6.63 &  SDSS       & 174    &  SDSS      & -21.42 &      SDSS            \\
        NGC 4945$^\dagger$                         &    563   & 1.7 & Henkel05             &        ... & ...  &  ...        & 134    &  Oliva95   & -21.07 &      DeV91           \\
        NGC 4968                                   &   2957   & ... & ...                  &        ... & ...  &  ...        &    ... &  ...       &    ... &      ...             \\
        NGC 5194 (M 51A)                           &    463   &-0.2 & Henkel05             &    11358   & 8.44 &  Ho97       & 113    &  McElroy95 & -20.38 &      DeV91           \\
        NGC 5256S (MrK 0266)                       &   8353   & 1.5 & Henkel05             &     4902   & 5.09 &  SDSS       & 187    &  SDSS      & -21.06 &      SDSS            \\
        SBS 1344+527$^\dagger$                     &   8763   & ... & ...                  &     2675   & 3.97 &  SDSS       & 147    &  SDSS      & -20.04 &      SDSS            \\
        NGC 5347                                   &   2335   & 1.5 & Henkel05             &     4433   & 4.68 &  SDSS       & 90     &  SDSS      & -19.26 &      SDSS            \\
        2MASX J13553592+0553050                    &  11776   & ... & ...                  &      285.9 & 6.08 &  SDSS       & 127    &  SDSS      & -20.03 &      SDSS            \\
        (IRAS 13530+0607 )                         &          &     &                      &            &      &             &        &            &        &                      \\
        MCG +11-17-010                             &   9456   & ... & ...                  &      136.9 & 5.08 &  SDSS       & 129    &  SDSS      & -20.06 &      SDSS            \\
        ESO 446-G018                               &   4771   & ... & ...                  &        ... & ...  &  ...        &    ... &  ...       &    ... &      ...             \\
        NGC 5495$^\dagger$                         &   6737   & 2.3 & Kondratko06          &        ... & ...  &  ...        &    ... &  ...       & -21.78 &      DeV91           \\
        Circinus$^\dagger$                         &    434   & 1.3 & Henkel05             &     30180  & 19.1 &  Bassani99  & 168    &  Oliva95   &    ... &      ...             \\
        NGC 5506 (MrK 1376)                        &   1853   & 1.7 & Henkel05             &    45744   & 7.2  &  Bassani99  & 180    &  Oliva99   & -19.50 &      SDSS            \\
        NGC 5643                                   &   1199   & 1.4 & Henkel05             &    74804   & 6.4  &  Bassani99  &    ... &  ...       & -21.20 &      DeV91           \\
                                                   &          & 1.3 & KGM06                &            &      &             &        &            &        &                      \\
        NGC 5691                                   &   1870   & ... & ...                  &        ... & ...  &  ...        &    ... &  ...       & -19.69 &      SDSS            \\
        NGC 5728$^\dagger$                         &   2804   & 1.9 & Henkel05             &    68000   & ...  &  Whittle92  & 209    &  McElroy95 & -21.18 &      DeV91           \\
        CGCG 164-019                               &   8963   & ... & ...                  &        ... & ...  &  ...        &    ... &  ...       & -20.37 &      SDSS            \\
        UGC 9618$^\dagger$                         &  10103   & ... & ...                  &        ... & ...  &  ...        &    ... &  ...       & -21.16 &      SDSS            \\
        MrK 834 (UGC 9639)$^\dagger$               &  10802   & ... & ...                  &     1554   & 6.52 &  SDSS       & 177    &  SDSS      & -21.79 &      SDSS            \\
        NGC 5793$^\dagger$                         &   3491   & 2.0 & Henkel05             &        ... & ...  &  ...        &    ... &  ...       & -19.75 &      DeV91           \\
        2MASX J15201964+5253560                    &  11166   & ... & ...                  &      203.3 & 5.04 &  SDSS       & 107    &  SDSS      & -20.23 &      SDSS            \\
        2MASX J16070391+0106296                    &   8216   & ... & ...                  &      165.5 & 5.4  &  SDSS       & 126    &  SDSS      & -19.47 &      SDSS            \\
        IRAS 16288+3929                            &   9161   & ... & ...                  &     3595   & 4.37 &  SDSS       & 142    &  SDSS      & -20.33 &      SDSS            \\
        CGCG 168-018                               &  11015   & ... & ...                  &      805.8 & 4.92 &  SDSS       & 118    &  SDSS      & -20.00 &      SDSS            \\
        NGC 6240 (IC 4625)                         &   7339   & 1.6 & Henkel05             &     795    & 17.2 &  Bassani99  & 200    &  Gerssen04 & -21.76 &      DeV91           \\
        NGC 6264$^\dagger$                         &  10177   & ... & ...                  &     5383   & 3.53 &  SDSS       & 149    &  SDSS      & -20.91 &      SDSS            \\
        2MFGC 13581                                &  10290   & ... & ...                  &     5947   & 3.79 &  SDSS       & 122    &  SDSS      & -20.44 &      SDSS            \\
        (2MASX J16581548+3923294)$^\dagger$        &          &     &                      &            &      &             &        &            &        &                      \\
        2MASX J17101815+1344058                    &   9448   & ... & ...                  &      ...   & ...  &  ...        &    ... &  ...       &    ... &      ...             \\
        (IRAS F17080+1347)                         &          &     &                      &            &      &             &        &            &        &                      \\
        NGC 6323$^\dagger$                         &   7772   & 2.7 & Henkel05             &      ...   & ...  &  ...        &    ... &  ...       & -20.58 &      DeV91           \\
        NGC 6300                                   &   1109   & 0.5 & Henkel05             &     1349   & 7.44 &  Polletta96 &    ... &  ...       & -20.53 &      DeV91           \\
                                                   &          & 0.34& KGM06                &            &      &             &        &            &        &                      \\
        ESO 103-G035                               &   3983   & 2.6 & Henkel05             &    12589   & 6.31 &  Polletta96 & 114    &  Gu06      & -19.51 &      DeV91           \\
        2MASX J19393889-0124328                    &   6622   & 2.2 & Henkel05             &      ...   & ...  &  ...        &    ... &  ...       &    ... &      ...             \\
        (IRAS F19370-0131)                         &          &     &                      &            &      &             &        &            &        &                      \\
        3C 403$^\dagger$                           &  17688   & 3.3 & Henkel05             &      ...   & ...  &  ...        &    ... &  ...       &    ... &      ...             \\
        NGC 6926$^\dagger$                         &   5880   & 2.7 & Henkel05             &      ...   & ...  &  ...        &    ... &  ...       & -22.29 &      DeV91           \\
        UGC 11685                                  &   5872   & ... & ...                  &      ...   & ...  &  ...        &    ... &  ...       &    ... &      ...             \\
        IC 1361                                    &   3962   & ... & ...                  &      ...   & ...  &  ...        &    ... &  ...       &    ... &      ...             \\
        AM 2158-380 NED02                          &   9983   & 2.7 & Kondratko06          &      ...   & ...  &  ...        &    ... &  ...       & -21.24 &      DeV91           \\
        2MASX J22291248-1810470$^\dagger$          &   7520   & 3.8 & Henkel05             &      ...   & ...  &  ...        &    ... &  ...       &    ... &      ...             \\
        (IRAS F22265-1826)                         &          &     &                      &            &      &             &        &            &        &                      \\
        NGC 7479                                   &   2381   & 1.3 & BG08                 &     1103   & 9.29 &  Ho97       & 155    &  Ho09      & -21.64 &      DeV91           \\
        IC 1481                                    &   6118   & 2.5 & Henkel05             &      ...   & ...  &  ...        &    ... &  ...       & -20.77 &      DeV91           \\
        CGCG 498-038                               &   9240   & ... & ...                  &      ...   & ...  &  ...        &    ... &  ...       &    ... &      ...             \\

\enddata
\tablenotetext{1}{Heliocentric systemic velocity, in $\kms$.}
\tablenotetext{2}{Total maser luminosity assuming isotropic emission of radiation, 
defined as $\lmaser = 0.023 \times \int S(V)dV \times D^2$,
where \lmaser~is in \Lsun, $S(V)$ is the flux (in Jy) at velocity $V$ (in $\kms$),
and $D$ is luminosity distance (in Mpc).  These values were calculated based on $\hh=75~\kms$ and
We convert values into those with $\hh=70~\kms$ in the analysis.}
\tablenotetext{3}{References -- Henkel05: \citet{henkel05a}; Kondratko05: \citet{kondratko05a}; 
Kondratko06: \citet{kondratko06b}; KGM06: \citet{kondratko06a}; Zhang06: \citet{zhang06a}; 
BG08: \citet{braatz08a}; Castangia08: \citet{castangia08a}; Surcis09: \citet{surcis09a}.}
\tablenotetext{4}{Observed flux of \oiiilam{}, in $10^{-17}~\ergscm$.}
\tablenotetext{5}{Ratio of observed flux of \Ha~to that of \Hb.}
\tablenotetext{6}{References -- DD88: \citet{dahari88a}; Whittle92: \citet{whittle92a}; 
Polletta96: \citet{polletta96a}; Ho97: \citet{ho97a}; Bassani99: \citet{bassani99a}.
In the compilation, we first prefer the homogeneous samples from SDSS and Ho97.
We then choose the most recent values for the rest of the sample.}
\tablenotetext{7}{Velocity Dispersion $\sigma$, in $\kms$.}
\tablenotetext{8}{References -- 
McElroy95: \citet{mcelroy95a}; NW95: \citet{nelson95a}; Oliva95: \citet{oliva95a}; 
Oliva99: \citet{oliva99a}; Wegner03: \citet{wegner03a}; Gerssen04: \citet{gerssen04a}; Gu06: \citet{gu06a}; GH06: \citet{greene06a};
Ho09: \citet{ho09a}.}
\tablenotetext{9}{$B$ band absolute magnitude.}
\tablenotetext{10}{References -- DeV91: \citet{devaucouleurs91a}. Magnitudes are corrected for foreground Galatic extinction \citep{schlegel98a}.}
\tablenotetext{a}{The merging system Arp 299 consists of two galaxies, NGC 3690 and IC 694. 
We assume the maser is associated with NGC 3690.}
\tablenotetext{b}{Assumed to be $3$ in the analysis.}
\tablenotetext{$\dagger$}{~Likely disk systems.}
\label{table:data}
\end{deluxetable*}
\clearpage

\end{document}